\documentclass[journal]{IEEEtran}

\usepackage{amsmath,amsfonts}
\usepackage{array}
\usepackage{textcomp}
\usepackage{stfloats}
\usepackage{url}
\usepackage{verbatim}
\usepackage{graphicx}
\usepackage{cite}

\usepackage{amssymb}
\usepackage{bm}
\usepackage{multirow}
\usepackage{booktabs}
\usepackage{lipsum}
\usepackage{enumerate}
\usepackage{multicol}
\usepackage{multirow}
\usepackage{subfigure}
\usepackage{stfloats}
\usepackage{color}

\usepackage[ruled,linesnumbered]{algorithm2e}

\usepackage{halloweenmath}
\allowdisplaybreaks[4]

\usepackage{amsthm}
\usepackage{ulem}
\newtheoremstyle{mytheostyle}{1pt}{1pt}{}{\parindent}{\bfseries}{\textnormal{\textit{:}}}{.5em}{\textit{\uline{\thmname{#1}\thmnumber{ #2}}}\thmnote{ \textnormal{\textit{(#3)}}}}
\theoremstyle{mytheostyle}

\newtheorem{remark}{Remark}

\newtheorem{proposition}{Proposition}

\newtheorem{theorem}{Theorem}
\newtheorem{insight}{Insight}
\newtheorem{assumption}{Assumption}

\newcommand{\mathbm}[1]{{\bm{#1}}}

\begin{document}

\title{Massive MIMO-DFRC System With\\1-Bit ADCs/DACs}

\author{Bowen Wang,~\IEEEmembership{Graduate Student Member,~IEEE},
        % <-this % stops a space
        Hongyu Li,~\IEEEmembership{Member,~IEEE},\\
        % <-this % stops a space
        Bin Liao,~\IEEEmembership{Senior Member,~IEEE},
        % <-this % stops a space
        and Ziyang Cheng,~\IEEEmembership{Senior Member,~IEEE}\vspace{-2em}
\thanks{This manuscript was posted as a preprint on arXiv \cite{wang2024massive_arXiv}.
Parts of this work were presented at the IEEE 2023 GLOBECOM Workshops \cite{wang2023joint}.
\textit{(Corresponding author: Ziyang Cheng)}
}
\thanks{Bowen Wang is with the School of Information and Communication Engineering, University of Electronic Science and Technology of China, Chengdu, China, and is with the Department of Engineering, King’s College London, London, WC2R 2LS, UK. (e-mail: bwwang@ieee.org).}
\thanks{Ziyang Cheng is with the School of Information and Communication Engineering, University of Electronic Science and Technology of China, Chengdu, China. (e-mail: zycheng@uestc.edu.cn).}% <-this % stops a space
\thanks{Hongyu Li is with the Internet of Things Thrust, The Hong Kong University of Science and Technology (Guangzhou), Guangzhou 511400, China (e-mail: hongyuli@hkust-gz.edu.cn).}
\thanks{Bin Liao is with the Guangdong Key Laboratory of Intelligent Information Processing, College of Electronics and Information Engineering, Shenzhen University, Shenzhen 518060, China. (e-mail: binliao@szu.edu.cn).}
}

\maketitle

\begin{abstract}
Massive multiple-input multiple-output (MIMO) dual-functional radar-communication (DFRC) has emerged as a key enabler for next-generation aerospace, wireless, and electronic systems.
To enhance energy efficiency and reduce hardware cost, we consider a hardware-friendly architecture, termed 1BitDFRC, which employs 1-bit digital-to-analog converters (DACs) at the DFRC transmitter and 1-bit analog-to-digital converters (ADCs) at the radar sensing receiver. 
Two joint transceiver designs are considered: \textit{1):} quality-of-service constrained scheme, and \textit{2):} quality-of-detection constrained scheme. 
These problems are non-trivial due to the 1-bit quantization, which induces continuous-to-discrete-value transformations and thus complicates both communication and radar sensing performance analysis. 
To gain some insights on the impact of such quantization on the performance of communication and radar sensing, we first characterize radar detection after 1-bit ADC quantization and analyze the communication bit error rate. 
Leveraging these insights, we define DFRC-oriented performance metrics that enable tractable reformulations of the design problems, which are efficiently solved via majorization–minimization and integer linear programming. 
The proposed analysis and design framework are extended to the multi-target case, showcasing the generality of the proposed methods.
Numerical results validate the analysis and demonstrate that 1BitDFRC achieves an advantageous trade-off between sensing/communication performance and energy efficiency compared with DFRC equipped with only 1-bit DACs, only 1-bit ADCs, or ideal infinite-resolution transceivers.
\end{abstract}


\vspace{-1em}
\section{Introduction}

\IEEEPARstart{W}{ireless}
communication and radar are two major applications of electromagnetic (EM) waves, where radar utilizes EM waves to detect, estimate, and track targets, while wireless communication uses EM waves to convey information. 
Although wireless communication and radar share many similarities, they have been developed independently over the past few decades \cite{liu2023seventy}.
Recently, future 6G networks and next-generation defense systems are envisioned to simultaneously support wireless communication and environment-aware radar sensing \cite{zhang2021overview}.
This increasing demand has driven research on dual-functional radar-communications (DFRC), where a common hardware platform and waveform support both radar sensing and communication, thereby enabling joint exploration of the two functionalities \cite{liu2020joint}.
With its advantages in spectrum sharing, hardware efficiency, and integration gain, DFRC has been recognized as a pivotal technology for next-generation wireless networks and defense systems \cite{xiong2024torch}.

\vspace{-1em} \subsection{Related Works}

In general, DFRC design methodologies can be classified into three categories: communication centric design (CCD), radar centric design (RCD), and dual-function waveform design (DFWD).

\textit{1) Communication-Centric Design:}
The CCD leverages classical communication waveforms and standards, introducing slight modifications to enable radar sensing as a secondary functionality \cite{feng2011received,jiang2008ieee,kumari2017ieee,li2024frame}.
For example, the authors in \cite{feng2011received} utilize the Wi-Fi standard to achieve indoor sensing, where indoor sensing is realized by carefully modifying the Wi-Fi standard to extract indoor environment information.
Besides, the authors in \cite{jiang2008ieee} discuss using the IEEE 802.11p standard to support vehicular environment sensing.
In addition to modifying IEEE 802.11p,  modifications to the IEEE 802.11ad standard have been presented in \cite{kumari2017ieee} to enhance sensing capabilities in vehicular environment.

\textit{2) Radar-Centric Design:} 
Similar to CCD, the RCD aims to embedded communication symbols into existing radar waveforms to achieve the communication as a secondary functionality \cite{nowak2016co,wu2021frequency,ahmed2018dual}.
For instance, \cite{nowak2016co} embeds binary phase shift keying (BPSK) symbols into linear frequency modulation (LFM) waveforms.
As a further step, it is proposed in \cite{wu2021frequency} to embed PSK symbols into a frequency-hopping (FH) waveform, where the embedding strategy of communication symbols is carefully studied.
To further enhance the communication rate,  an innovative approach that embeds quadrature amplitude modulation signals into the sidelobes of radar waveforms has been developed in \cite{ahmed2018dual}.

\textit{3) Dual-Function Waveform Design:} 
The CCD and RCD prioritize either communication or radar sensing capabilities in the DFRC system, treating the other functionality as secondary, which leads to performance losses on the secondary functionality.
To overcome this drawback and achieve enhanced and balanced DFRC performance, the DFWD approaches have been investigated in \cite{meng2023network,liao2024faster,liu2020TSP,liu2021dual,liu2021cramer,Guo2023TSP,wei2023waveform,wang2023relative}. 
Specifically, DFWD focuses on designing waveforms (or beamformers) that simultaneously optimize radar sensing and communication performance. 
Typical radar metrics encompass signal-to-interference-plus-noise ratio (SINR) \cite{meng2023network,liao2024faster}, beampattern mean squared error (MSE) \cite{liu2020TSP,liu2021dual}, Cramér–Rao bound (CRB) \cite{liu2021cramer,Guo2023TSP}, and mutual information \cite{wei2023waveform,wang2023relative}. 
On the other hand, communication performance is often evaluated in terms of achievable rate \cite{wei2023waveform}, user SINR \cite{liu2020TSP}, minimum MSE (MMSE) \cite{wang2023relative}, and constructive interference (CI) \cite{liu2021dual}. 
The resulting DFWD formulations are typically addressed via numerical optimization or machine learning techniques.

\vspace{-0.5em} \subsection{Motivations and Contributions}

In forthcoming 6G and defense systems, massive multiple-input multiple-output (MIMO) is considered a key technology \cite{Sun2025JSAC,Sun2024TWC} that enables orders-of-magnitude improvements in communication throughput and supports millimeter-level sensing.
Therefore, synergizing DFRC with massive MIMO technology has become increasingly promising for achieving high-throughput communication and higher-precision radar sensing.
However, directly extending the aforementioned approaches \cite{meng2023network,liao2024faster,liu2020TSP,liu2021dual,liu2021cramer,Guo2023TSP,wei2023waveform,wang2023relative} to massive MIMO DFRC systems presents significant challenges.
Specifically, previous studies assume DFRC is equipped with a fully-digital beamforming (FD-BF) architecture, where each antenna is associated with a dedicated RF chain.
The deployment of a large antenna array in massive MIMO DFRC systems necessitates a corresponding number of RF chains, thereby incurring prohibitive power consumption and hardware expense.
Therefore, it is imperative and remains unexplored to investigate energy and hardware-efficient solutions for massive MIMO-DFRC systems.

To overcome this challenge, low-resolution massive MIMO (LowRes-MIMO) architecture has been introduced, where each RF chain is equipped with few-bit ADCs/DACs.
LowRes-MIMO was first introduced in communication areas \cite{zhang2018low,castaneda20171,li20211,mo2017channel,Shao2019TSP,Wu2023CI,meng2018generalized,Nghi2020TWC,Nghi2025Physical} and has received significant research attention. 
Subsequently, it has been extended to radar applications \cite{cheng2023relative,xi2020gridless,xiao2022one,deng2022receive,shang2024mixed,Liao2019MUSIC,Sedighi2021TSP}.
Given the successful application of LowRes-MIMO in both communication and radar areas, researchers propose achieving massive MIMO DFRC by employing LowRes-MIMO architecture \cite{cheng2021transmit,yu2022precoding,Lin2025TCOM}.
Specifically, prior work conducted in \cite{cheng2021transmit} designs an DFRC system with 1-bit DACs by minimizing the CRB while guaranteeing the MMSE in communications.
Subsequently, \cite{yu2022precoding} develops a 1-bit DAC DFRC precoding scheme that minimizes a weighted combination of radar beampattern MSE and communication MMSE.

Although \cite{cheng2021transmit,yu2022precoding,Lin2025TCOM} achieve satisfactory DFRC performance, they have the following  limitations:

\textit{First}, the aforementioned works \cite{cheng2021transmit,yu2022precoding} focus exclusively on DFRC with 1-bit DACs at the transmitter, leaving the scenario with 1-bit ADCs/DACs largely unexplored. 
The most relevant work \cite{Lin2025TCOM} evaluates radar performance in DFRC with 1-bit ADCs using Bussgang theory \cite{papoulis1965random}. 
However, Bussgang theory \cite{papoulis1965random} is a statistical approximation technique, and its accuracy under 1-bit ADCs quantization, especially for sensing tasks, remains questionable and requires further validation. 
On the other hand, in radar-only settings, our prior work \cite{deng2022receive} formulates a detector under idealized conditions by assuming that the radar signal after receive filtering is independent and identically distributed, which neglects the statistical dependencies introduced by the receive filter.
These observations raise a critical question: ``\textit{Can we evaluate radar performance in a more general and practical setting—without relying on Bussgang or overly strict statistical assumptions?}"

\textit{Second}, the aforementioned works \cite{cheng2021transmit,yu2022precoding,Lin2025TCOM} evaluate communication performance using the MMSE criterion, which is effective in mitigating multi-user interference (MUI) under high-resolution transceiver settings. 
However, when the DFRC transmitter is constrained by 1-bit DACs, the ability to suppress MUI becomes severely limited due to coarse quantization and reduced design flexibility. 
Recent advances in symbol error probability (SEP) \cite{li2013distributed,salem2021error,shao2019framework} analysis and constructive interference (CI) based symbol level precoding (SLP) method \cite{li2020interference,li2018massive,li20211,Wu2023CI} have offered new design perspectives that potentially provide more design flexibility. 
These considerations lead to another question: ``\textit{Is MMSE still an effective performance metric for communication in DFRC systems with 1-bit DACs? Can SEP/SLP-based analysis offer a more practical and design-relevant alternative?}"

\textit{Third}, the aforementioned works \cite{cheng2021transmit,yu2022precoding,Lin2025TCOM} design DFRC systems with 1-bit DACs using alternative optimization. 
For instance, \cite{Lin2025TCOM} formulates a simplified weighted-sum problem under only 1-bit DAC constraints, which eases optimization. 
Moreover, \cite{cheng2021transmit,yu2022precoding} introduce strong penalty terms to approximate the non-convex 1-bit constraint, leading to slow convergence and lacking optimality guarantees. 
To better capture real-world application scenarios, this paper considers general and realistic quality of service (QoS)/quality of detection (QoD)-constrained designs, where the increased complexity calls for more efficient and theoretically guaranteed approaches.
These observations motivate the question:
\textit{How can we design optimization algorithms for 1-bit DFRC systems that effectively address practical QoS/QoD constraints while ensuring both efficiency and convergence guarantees?}

To answer the above three questions, this paper proposes a 1-bit ADCs/DACs empowered DFRC system, termed as 1BitDFRC, and thoroughly studies the DFRC performance based on theoretical analysis and optimizations. 
The main contributions of this paper are outlined below:

\textit{First, 1BitDFRC Performance Analysis.}
Under practical and general statistical conditions, we comprehensively analyze the detection performance of 1BitDFRC, revealing a performance loss of only 1.96 dB caused by 1-bit ADC quantization in 1BitDFRC systems with massive MIMO.
The analysis offers valuable guidance for the design and optimization of 1BitDFRC systems.
Additionally, the SEP performance bound of the considered 1BitDFRC is established, to evaluate communication bit error rate (BER) performance, which indicates that ensuring the received signal lies within the CI-margin is sufficient to guarantee the communication BER performance.

{\textit{Second, Joint Transceiver 1BitDFRC Design.}}
Two 1BitDFRC designs are proposed in this paper: the QoS-constrained 1BitDFRC design and the QoD-constrained 1BitDFRC design. 
Specifically, the QoS-constrained 1BitDFRC design aims to maximize radar detection performance while adhering to communication QoS constraints. 
Guided by insightful performance analysis, the QoS-constrained 1BitDFRC design problem is reformulated into a more tractable form and subsequently solved using a proposed method that combines integer linear programming (ILP) and majorization-minimization (MM).
Additionally, we extend the proposed QoS-constrained 1BitDFRC design to a QoD-constrained 1BitDFRC design, where radar detection performance is constrained while communication QoS is maximized.

{\textit{Third, Extension to Multi-Target Scenario.}}
The proposed detection performance analysis of 1BitDFRC is extended to the multi-target scenario, showcasing the generality of our analytical framework. 
Furthermore, a fairness-aware multi-target QoS-constrained 1BitDFRC design problem is formulated and solved by extending the method developed in above \textit{Second} point.

{\textit{Fourth, Performance Validation and Comparison.}}
Comprehensive simulation results are presented to verify the analytical findings on 1BitDFRC. 
In addition, comparisons with other DFRC configurations—featuring only 1-bit ADCs or 1-bit DACs—are carried out to assess their practicality in massive MIMO systems. 
The results demonstrate that the proposed 1BitDFRC attains an effective balance between DFRC performance and energy efficiency.

\vspace{-0.5em} \subsection{Notations}
Vectors and matrices are denoted by lowercase boldface letters (e.g., $\mathbm{a}$) and uppercase boldface letters (e.g., $\mathbm{A}$), respectively.
$\mathbb{C}^n$ and $\mathbb{C}^{m \times n}$ denote the $n$-dimensional complex vector space and the $m \times n$ complex matrix space, respectively.
The operators $(\cdot)^T$, $(\cdot)^H$, and $(\cdot)^{-1}$ represent transpose, conjugate transpose, and matrix inversion, respectively.
$\Re \{ \cdot \}$ and $\Im \{ \cdot \}$ denote the real and imaginary parts of a complex quantity.
$\| \cdot \|_F$ and $| \cdot |$ denote the Frobenius norm and the absolute value, respectively.
$\mathrm{Tr}(\cdot)$ denotes the trace of a matrix, and $\mathrm{Diag}(\cdot)$ denotes a diagonal matrix constructed from its arguments.

\section{Signal Model}\label{Sec:2}
In this section, models for the 1BitDFRC system are introduced and problem statements for the 1BitDFRC designs are presented.

\begin{figure}[t]
	\centering
	\includegraphics[width=0.92\linewidth]{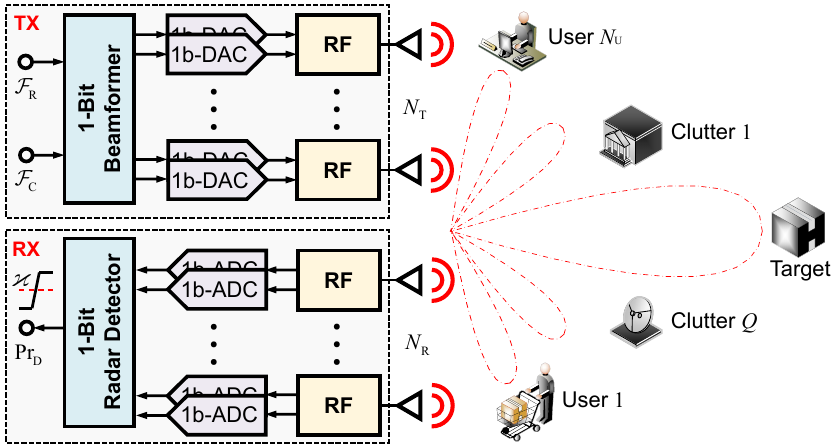}
	\caption{Architecture of the considered 1BitDFRC.}
	\label{fig:sys}
	\vspace{-1em}
\end{figure}

\vspace{-0.5em} \subsection{1BitDFRC Model}

As shown in Fig. \ref{fig:sys}, we consider a 1BitDFRC scenario consisting of a dual-function BS (DFBS), a target, $Q$ clutter sources, and $N_\mathrm{U}$ single-antenna user equipments (UEs).
The DFBS simultaneously transmits a dual-function waveform to probe the target of interest and provide communication services to the UEs.
The DFBS receives echo signals, aiming to detect the targets while suppressing clutter.
To reduce hardware costs and power consumption, the DFBS employs 1-bit DACs at its transmitter and 1-bit ADCs at its receiver, with the transmitter and receiver assumed to be co-located.

In the 1BitDFRC transmitter, according to the radar and communication functions, the communication signal $\mathbm{s} = [s_1 , \cdots , s_{N_\mathrm{U}}]^T$ is precoded into an unquantized transmit signal $\mathbm{x}_\mathrm{o}$ via a beamformer $\mathcal{B}$, i.e., $\mathbm{x}_\mathrm{o} = \mathcal{B}(\mathbm{s}) \in\mathbb{C}^{N_\mathrm{T}}$.
Then, the baseband signal $\mathbm{x}_\mathrm{o}$ is quantized by the 1-bit DACs, and the transmit signal is expressed as
\begin{equation}\label{eq:1}
	\mathbm{x} = \mathcal{Q}_{\text{DAC}}^{\text{1-Bit}}\left( \mathbm{x}_{\mathrm{o}} \right) =
	\mathcal{Q}_{\text{DAC}}^{\text{1-Bit}}\left( \mathcal{B}\left( \mathbm{s} \right) \right) 
	\in \mathcal{X}_{\text{DAC}}^{\text{1-Bit}} ,
\end{equation}
where 
\begin{equation}\label{eq:1_bit}
	\mathcal{Q}_{\mathrm{DAC}}^{\mathrm{1\text{-}bit}}(z)
	\triangleq
	\sqrt{\frac{E}{2N_{\mathrm T}}}
	\left( \mathrm{sign}[\Re(z)] + \jmath\,\mathrm{sign}[\Im(z)] \right) 
\end{equation}
represents the 1-bit quantization operator.
The output of $\mathcal{Q}_{\text{DAC}}^{\text{1-Bit}}(\cdot)$ falls within the quantization alphabet $\mathcal{X}_{\text{DAC}}^{\text{1-Bit}} = \left\{ \pm \sqrt{\frac{E}{2N_\mathrm{T}}} , \pm \jmath \sqrt{\frac{E}{2N_\mathrm{T}}} \right\}$, where $E$ denotes the transmit power budget.

The aforementioned workflow implies that the transmit symbols in $\mathbm{s}$ go through a two-stage transformation and end up with being mapped to elements in the 1-bit alphabet $\mathcal{X}_{\text{DAC}}^{\text{1-Bit}}$ \cite{jacobsson2017quantized}. 
In this sense, there is no need to explicitly define the beamforming function $\mathcal{B}(\cdot)$ and we can directly optimize the 1-bit transmit signal $\bm{x}$ over the $\mathcal{X}_{\text{DAC}}^{\text{1-Bit}}$, as has been widely adopted in \cite{cheng2021transmit,yu2022precoding,Shao2019TSP,li20211,castaneda20171,jacobsson2017quantized}. 
A detailed solution to $\mathbm{x}$ will be provided  in Sections~\ref{Sec:4} and~\ref{Sec:5}.

In the following subsections, we elaborate on the radar and communication functions.

\vspace{-0.5em} \subsection{Radar Model}\label{Sec:2_B}

Consider a target of interest located at angle $\theta_0$ in the presence of $Q$ stationary clutter sources (e.g., trees, buildings) at angles $\theta_q, \forall q$. 
The corresponding echo signals received at the DFBS can be written as\footnote{In this paper, we assume that the proposed 1BitDFRC system operates in a pulse-based (time-division) DFRC mode \cite{li2024frame,liu2020joint,zhang2021overview}, where the system first transmits the DFRC signal and then switches to the receiving mode.
As a result, self-interference is inherently avoided and is therefore not considered.}
\begin{equation}\label{eq_3_New}
	\begin{aligned}
		\mathbm{r} = & \mathbm{r}_\mathrm{T} + \mathbm{r}_\mathrm{C} + \mathbm{n}_\mathrm{R} = \underbrace {\gamma _0\mathbm{G}_0 {\mathbm{x}} }_{\mathbm{r}_\mathrm{T}}
		+ \sum\limits_{q = 1}^Q \underbrace {{{\gamma _q}\mathbm{G}_q{\mathbm{x}}} }_{ \mathbm{r}_{\mathrm{C},q} } + \mathbm{n}_\mathrm{R} .
	\end{aligned}
\end{equation}
where $\mathbm{n}_\mathrm{R}$ is additive white Gaussian noise (AWGN) with $\mathbm{n}_\mathrm{R} \sim \mathcal{CN} \left( 0 , \sigma_\mathrm{R}^2 \mathbm{I}_{N_{\mathrm{R}}} \right)$.
$\gamma_0$ represents the complex amplitude of target with ${\gamma _0} \sim {\mathcal C}{\mathcal N}\left( {0,\varsigma _0^2} \right) $.
Similarly, ${\gamma _q}$ denotes the complex amplitude of clutter source $q$ with ${\gamma _q} \sim {\mathcal C}{\mathcal N}\left( {0,\varsigma _q^2} \right) , \forall q$, where the DFBS only knows the $\varsigma _q^2$ without accurate value of ${\gamma _q}$.
$\mathbm{G}_0 = {{\mathbm{g}}_\mathrm{R}}\left( {{\theta _0}} \right){\mathbm{g}}_\mathrm{T}^T\left( {{\theta _0}} \right)$ and
$\mathbm{G}_q = {{\mathbm{g}}_\mathrm{R}}\left( {{\theta _q}} \right){\mathbm{g}}_\mathrm{T}^T\left( {{\theta _q}} \right)$ respectively denote the equivalent radar channel for target and clutter $q$.
$\mathbm{g}_\mathrm{T}(\theta)$ and $\mathbm{g}_\mathrm{R}(\theta)$ denotes steering vectors, given by
$\mathbm{g}_\mathrm{T}(\theta) = \left[ 1 , e^{-\jmath \pi \sin {\theta}}, \cdots, e^{-\jmath \pi (N_{\mathrm{T}}-1)\sin {\theta}} \right] ^T / {\sqrt{N_\mathrm{T}}}$ and $\mathbm{g}_\mathrm{R}(\theta) = \left[ 1 , e^{-\jmath \pi \sin {\theta}}, \cdots, e^{-\jmath \pi (N_{\mathrm{R}}-1)\sin {\theta}} \right] ^T / {\sqrt{N_\mathrm{R}}}$.
The received signal $\mathbm{r}$ is then quantized by 1-bit ADCs, the quantized output signal is given by
\begin{equation}
	\widetilde{\mathbm{r}} = \mathcal{Q}_{\text{ADC}}^{\text{1-Bit}} \left( \mathbm{r} \right) = \mathcal{Q}_{\text{ADC}}^{\text{1-Bit}} \left( \mathbm{r}_\mathrm{T} + \mathbm{r}_\mathrm{C} + \mathbm{n}_\mathrm{R} \right) ,
\end{equation}
where $\mathcal{Q}_{\text{ADC}}^{\text{1-Bit}} (\cdot)$ is the 1-bit quantization function, which is defined similarly to~\eqref{eq:1_bit} and can be mathematically described as $\mathcal{Q}_{\text{ADC}}^{\text{1-Bit}} (z) = \mathrm{sign}[\Re(z)] + \jmath \mathrm{sign}[\Im(z)]$.

After 1-bit ADC quantization, the output signal is processed by a receive filter ${\mathbm{f}}\in\mathbb{C}^{N_{\mathrm{R}}}$.
Then, binary hypothesis test for target detection is formulated as
\begin{equation}
	\left\{  
		\begin{array}{ll}
			\mathcal{H}_0: & z = {\mathbm{f}}^H\widetilde{\mathbm{r}} = {\mathbm{f}}^H \mathcal{Q}_{\text{ADC}}^{\text{1-Bit}}( \mathbm{r}_\mathrm{C} + \mathbm{n}_\mathrm{R} ) \\
			\mathcal{H}_1: & z = {\mathbm{f}}^H\widetilde{\mathbm{r}} = {\mathbm{f}}^H \mathcal{Q}_{\text{ADC}}^{\text{1-Bit}}( \mathbm{r}_\mathrm{T} + \mathbm{r}_\mathrm{C} + \mathbm{n}_\mathrm{R} )
		\end{array}
	\right. .
	\label{eq:6}
\end{equation}
For notational brevity, we define $\bm{r}_i$ as the received signal from the target and clutters under hypothesis $\mathcal{H}_i$.
Specifically, under $\mathcal{H}_1$, we have $\bm{r}_1 = \bm{r}_\mathrm{T} + \bm{r}_\mathrm{C}$, while under $\mathcal{H}_0$, the received signal reduces to $\bm{r}_0 = \bm{r}_\mathrm{C}$.

Finally, based on \eqref{eq:6}, the generalized likelihood ratio test (GLRT) for radar detection is derived as:
\begin{equation}\label{eq:7}
	 \frac{{{\mathrm{Pr}}\left( {z|{{\mathcal H}_1}} \right)}}{{{\mathrm{Pr}}\left( {z|{{\mathcal H}_0}} \right)}}\mathop  \gtrless \limits_{{{\mathcal H}_0}}^{{{\mathcal H}_1}} \varkappa ,
\end{equation}
where $\varkappa$ is the detection threshold and ${{\mathrm{Pr}}\left( {z|{{\mathcal H}_i}} \right)} , i  \in \{0 , 1\}$ represents the probability density function (PDF) under $\mathcal{H}_i$.

\vspace{-0.5em} \subsection{Communication Model}
Beyond radar sensing, the DFBS transmit waveform conveys communication symbols to the UEs. 
The received signal at UE-$u$ can thus be written as:
\begin{equation}\label{eq:8}
	\begin{aligned}
		y_u & = \mathbm{h}_u^H \mathbm{x} + n_{\mathrm{C},u} ,
	\end{aligned}
\end{equation}
where $\mathbm{h}_u$ is the communication channel from DFBS to UE-$u$, and ${n_{\mathrm{C},u}}$ is the AWGN with ${n_{\mathrm{C},u}} \sim {\mathcal {CN}}\left( {0,\sigma _{\mathrm{C},u}^2} \right)$.
Given that UEs employ only a single antenna, the hardware cost and power consumption are modest, hence, it is reasonable to assume that they utilize high-resolution ADCs.

In this paper, we assume the communication symbol $s_u$ for UE-$u$ is selected from an $\mathcal{M}$-PSK alphabet with $\mathcal{S}_\mathcal{M} = \{ e^{\jmath \frac{2\pi (m-1)}{\mathcal{M}}}, m = 1 , \cdots , \mathcal{M} \}$.
Therefore, UE-$u$ strives to detect the desired symbol $s_u$ from the $y_u$.
Accordingly, the BER for the 1BitDFRC system can be measured via the pairwise SEP, which is defined as
\begin{equation}\label{eq:9}
	\mathrm{SEP}_u = {\mathrm{Pr}}\left( {{y_u} \nrightarrow {s_u}} \right) \triangleq {\mathrm{Pr}}\left( {{y_u} \notin {\mathcal R}\left( {{s_u}} \right)} \right) ,
\end{equation}
where ${{\mathcal R}\left( {{s_u}} \right)}$ denotes the detection region of ${{s_u}}$.

Note that as long as the pairwise SEP is relatively low, the BER of the considered 1BitDFRC system will be minimal, thereby resulting in high-quality communication.

\vspace{-0.5em} \subsection{Problem Statement}
Achieving high-quality DFRC requires enhancing both radar detection capability and communication BER performance in the proposed 1BitDFRC architecture.
Towards this end, two 1BitDFRC designs are proposed, i.e., the QoS-constrained 1BitDFRC design and the QoD-constrained 1BitDFRC design.

\subsubsection{QoS-Constrained 1BitDFRC Design}
In the first design, we aim to maximize the probability of detection $\mathrm{Pr}_\mathrm{D}$ while ensuring constant probability of false alarm $\mathrm{Pr}_\mathrm{FA}$ and and maintaining the BER (SEP) below a predefined threshold, as given by
\begin{equation}\label{eq:QoS}
	\begin{aligned}
		\mathop {\max }\limits_{\mathbm{x} \in \mathcal{Q}_{\text{DAC}}^{\text{1-Bit}},{{\mathbm{f}}}} \; \mathrm{Pr}_\mathrm{D} \quad
		{\text{s}}.{\text{t}}. \; \mathrm{Pr}_\mathrm{FA} = \delta , \; \mathrm{SEP}_u \le \lambda_u , \forall u ,
	\end{aligned}
\end{equation}
where $\delta \in [0,1]$ is the desired probability of false alarm, and $\lambda_u$ is the required BER threshold.

\subsubsection{QoD-Constrained 1BitDFRC Design}
To facilitate a fair comparison of communication performance across users, the second design criterion minimizes the \emph{worst-user} communication SEP subject to the radar probability-of-detection $\mathrm{Pr}_{\mathrm D}$ requirement, which can be mathematically described as
\begin{equation}\label{eq:QoD}
	\begin{aligned}
		\mathop {\min }\limits_{\mathbm{x} \in \mathcal{Q}_{\text{DAC}}^{\text{1-Bit}},{{\mathbm{f}}}} \; \max_{u} \; \{ \mathrm{SEP}_u \} \quad
		{\text{s}}.{\text{t}}. \; \mathrm{Pr}_\mathrm{D} \ge \eta , \; \mathrm{Pr}_\mathrm{FA} = \delta , 
	\end{aligned}
\end{equation}
where $\eta \in [0,1]$ is the required probability of detection. 
The min--max objective in~\eqref{eq:QoD} is used to promote user fairness by balancing the SEP performance across users.

\begin{remark}
	The proposed designs in \eqref{eq:QoS} and \eqref{eq:QoD} are general formulations without specifying the explicit expressions of $\mathrm{Pr}_\mathrm{D}$ and $\mathrm{Pr}_\mathrm{FA}$, while they are extremely difficult to tackle due to the following twofold difficulties:
	\textit{1):} The proposed designs rely on the probabilities of detection and false alarm, together with the communication symbol error probability (SEP).
	However, given that we are considering 1BitDFRC, the closed-form expressions for $\mathrm{Pr}_\mathrm{D}$ and $\mathrm{Pr}_\mathrm{FA}$ have not yet been determined.
	Furthermore, developing effective methods to measure and enhance communication SEP also warrants significant research.
	\textit{2):} The proposed designs restrict $\mathbm{x}$ to belong to $\mathcal{X}_{\text{DAC}}^{\text{1-Bit}}$, resulting in non-convex problems with discrete constraints. 
	Moreover, the large number of antennas in massive MIMO systems significantly increases the complexity of solving these design problems.
\end{remark}

\begin{remark}\label{New_Rem_2}
	If the same metrics for $\mathrm{SEP}_u$, $\Pr_{\mathrm{D}}$, and $\Pr_{\mathrm{FA}}$ are specified, the design problems \eqref{eq:QoS} and \eqref{eq:QoD} will achieve the same radar sensing and communication performance bound \cite{Chen2021Pareto}. 
	This is because the key difference between \eqref{eq:QoS} and \eqref{eq:QoD} lies in how the objectives and constraints are assigned: \eqref{eq:QoS} maximizes radar detection performance $\Pr_{\mathrm{D}}$ subject to a communication QoS on $\mathrm{SEP}_u$, whereas \eqref{eq:QoD} minimizes $\mathrm{SEP}_u$ subject to a radar detection $\Pr_{\mathrm{D}}$ requirement. 
	Nevertheless, the two formulations are suited to different scenarios. 
	Specifically, \eqref{eq:QoS} is more appropriate when precise control over communication quality is required, whereas \eqref{eq:QoD} is preferable when tighter control over radar performance is the priority.
\end{remark}

To address the first difficulty, a comprehensive performance analysis, which leads to closed-form formulations of radar detection and communication metrics, is presented in Sec. \ref{Sec:3}. 
Following the provided design guidelines, to tackle the second difficulty, the problem reformulation and solutions for the proposed designs \eqref{eq:QoS}-\eqref{eq:QoD} are detailed in Sections \ref{Sec:4} and \ref{Sec:5}, respectively.

\section{Performance Analysis and Metrics}\label{Sec:3}
In this section, we first perform an analysis of radar detection after 1-bit ADC quantization to obtain design guidelines. 
Then, we focus on characterizing communication BER performance and derive the communication metric.

\vspace{-0.5em} \subsection{Radar Sensing Performance and Metric}\label{Sec:3-1}

In this subsection, we comprehensively analyze radar detection performance. 
Following this analysis, some insights are summarized.

\subsubsection{Radar Detection Performance Analysis}
The derivation of the detection performance comprises the following two steps:

\underline{Step 1}: PDF Derivation. 
Unlike the $\infty$-bit ADC case, the received signal is quantized by 1-bit ADCs, which makes the conditional PDF of $z$ under $\mathcal{H}_i$ difficult to derive.

To address this challenge, we exploit the following property of practical DFRC in the weak target detection regime.
\begin{assumption}[Weak Target Detection Regime]\label{property:1}
	Consider the element-wise signal-to-noise ratio (SNR) and clutter-to-noise ratio (CNR) at the $n$-th receive antenna, defined as
	${\rm SNR}_{\rm per}^n = \frac{|[\bm{r}_{\rm T} +\bm{r}_{\rm C}]_n|^2} {|[{\bm n}_{\rm R}]_n|^2} = \frac{|[\bm{r}_1 ]_n|^2} {|[{\bm n}_{\rm R}]_n|^2}$ 
	and 
	${\rm CNR}_{\rm per}^n = \frac{|[\bm{r}_{\rm C}]_n|^2} {|[{\bm n}_{\rm R}]_n|^2} = \frac{|[\bm{r}_0]_n|^2} {|[{\bm n}_{\rm R}]_n|^2}$,
	respectively.
	In the weak target detection scenario, both the target-plus-clutter return and the clutter return are below the receiver noise floor (low SNR/CNR property), i.e.,
	$|[\bm{r}_1]_n|^2 \ll \sigma_\mathrm{R}^2$ and $|[\bm{r}_0]_n|^2 \ll \sigma_\mathrm{R}^2$, for all $n$.
\end{assumption}

\textit{Assumption \ref{property:1}} characterizes a common operating regime in practical radar sensing, where the echo strength is limited by severe two-way propagation loss and the target RCS, especially when the target is distant from the transceiver.
As a result, the useful sensing return is typically weak, which motivates dedicated receive processing to reliably extract target information.

\begin{theorem}\label{the:1}
	Given the low SNR/CNR property, if $N_{\mathrm{R}} \to \infty$ such that the central limit theorem (CLT) holds, then the asymptotic PDF of $z = {\mathbm{f}}^H\widetilde{\mathbm{r}} = {\mathbm{f}}^H \mathcal{Q}_{\text{ADC}}^{\text{1-Bit}}( \mathbm{r}_i + \mathbm{n}_\mathrm{R} )$ can be formulated as
	\begin{equation}
		\text{Pr} \left( z | \mathcal{H}_i , \mathbm{\gamma}_i \right) = \frac{1}{2\pi \left\| {\mathbm{f}} \right\|_F^2 } \exp \Bigg( - \frac{ \big| z - \frac{2}{\sqrt{\pi \sigma_\mathrm{R}^2}} {\mathbm{f}}^H \mathbm{r}_i \big|^2 }{ 2 \left\| {\mathbm{f}} \right\|_F^2 } \Bigg) .
	\end{equation}
\end{theorem}
\begin{IEEEproof}
	See Appendix \ref{proof:1}.
\end{IEEEproof}

\begin{remark}
    In our prior work \cite{deng2022receive}, the asymptotic PDF of $z$ was derived under the assumption that the variables $\{[\mathbm{f}]_n^* [\widetilde{\mathbm{r}}]_n\}_{n=1}^{N_{\mathrm{R}}}$ are independent and identically distributed (i.i.d.), which is generally  unrealistic in practice. In the present paper, we relax this i.i.d. assumption and instead provide a more general result and formulation, the details of which are presented in Appendix \ref{proof:1}.
\end{remark}

Note that $ \mathrm{Pr}(z|\mathcal{H}_i) = \int {\mathrm{Pr}}\left( z | \mathcal{H}_i , \mathbm{\gamma}_i \right){\mathrm{Pr}}\left( \mathbm{\gamma}_i \right){\mathrm{d}}\mathbm{\gamma}_i $,
$\mathbm{\gamma}_0 \sim \mathcal{CN}( \mathbm{0} ,  \mathsf{Diag}( \varsigma _1^2 , \cdots , \varsigma _Q^2 ) )$, 
$\mathbm{\gamma}_1 \sim \mathcal{CN}( \mathbm{0} ,  \mathsf{Diag}( \varsigma _0^2 , \cdots , \varsigma _Q^2 ) )$, we have
\begin{equation}
	\mathrm{Pr}( z | \mathcal{H}_i )  = \frac{1}{\pi (2 \left\| {\mathbm{f}} \right\|_F^2 + \Pi_i)} \exp \left( - \frac{\left| z \right|^2}{ 2 \left\| {\mathbm{f}} \right\|_F^2 + \Pi_i} \right)  ,
\end{equation}
where 
$\Pi_1 = \frac{4\varsigma _0^2}{\pi \sigma_\mathrm{R}^2} |{\mathbm{f}}^H  {{\mathbm{G}}_0} {\mathbm{x}}|^2 + \sum_{q=1}^Q{ \frac{4\varsigma _q^2}{\pi \sigma_\mathrm{R}^2} |{\mathbm{f}}^H  {{\mathbm{G}}_q} {\mathbm{x}}|^2 } $ and 
$\Pi_0 =  \sum_{q=1}^Q{ \frac{4\varsigma _q^2}{\pi \sigma_\mathrm{R}^2} |{\mathbm{f}}^H  {{\mathbm{G}}_q} {\mathbm{x}}|^2 } $.

\underline{Step 2}: GLRT Derivation. With the PDFs at hand, based on above derivations and \eqref{eq:7}, the 1BitDFRC detector can be expressed as:
\begin{equation}
	z_\text{1Bit-G} = \left| z \right| \mathop  \gtrless \limits_{{{\mathcal H}_0}}^{{{\mathcal H}_1}} \varkappa_\text{1Bit-G} ,
\end{equation}
where $\varkappa_\text{1Bit-G}$ is detection threshold.

Since $\mathrm{Pr}( z | \mathcal{H}_0 )$ is a Gaussian variable with zero mean and variance $2 \left\| {\mathbm{f}} \right\|_F^2 + \Pi_0$, the $z_\text{1Bit-G} = \left| z \right|$ follows Rayleigh distribution.
Therefore, the probability of false-alarm $\mathrm{Pr}_\mathrm{FA}$ is given by
\begin{equation}
	\begin{aligned}
		\mathrm{Pr}_\mathrm{FA} & = \int_{\varkappa_\text{1Bit-G}}^{\infty}{ \mathrm{Pr}( z_\text{G} | \mathcal{H}_0 ) } \text{d}z_\text{G}  = \exp \left( - \frac{\varkappa_\text{1Bit-G}^2}{2 \left\| {\mathbm{f}} \right\|_F^2 + \Pi_0} \right) .
	\end{aligned}
	\label{eq:14}
\end{equation}

According to \eqref{eq:14}, for a given $\mathrm{Pr}_\mathrm{FA}$, the GLRT detection threshold is expressed as 
\begin{equation}
	\varkappa_\text{1Bit-G} = \sqrt{ - ( 2 \left\| {\mathbm{f}} \right\|_F^2 + \Pi_0 ) \text{ln} (\mathrm{Pr}_\mathrm{FA})  } .
	\label{eq:15}
\end{equation}

Similarly, since $\mathrm{Pr}( z | \mathcal{H}_1 )$ is a Gaussian variable with zero mean and variance $2 \left\| {\mathbm{f}} \right\|_F^2 + \Pi_1$, the $z_\text{1Bit-G} = \left| z \right|$ follows Rayleigh distribution.
The probability of detection $\mathrm{Pr}_\mathrm{D}$ with the $\varkappa_\text{1Bit-G}$ in \eqref{eq:15} is
\begin{equation}
	\begin{aligned}
		\mathrm{Pr}_\mathrm{D} & = \exp \Bigg( \frac{- \varkappa_\text{1Bit-G}^2}{2 \left\| {\mathbm{f}} \right\|_F^2 + \Pi_1} \Bigg)  = \exp\Bigg( \frac{\text{ln} (\mathrm{Pr}_\mathrm{FA})}{ 1 + \text{QSCNR}({\mathbm{f}} , \mathbm{x}) } \Bigg) ,
	\end{aligned}
	\label{eq:22}
\end{equation}
where $\text{QSCNR}({\mathbm{f}} , \mathbm{x})$ is given by
\begin{equation}
	\text{QSCNR}({\mathbm{f}} , \mathbm{x}) = \frac{2}{\pi} \frac{\mathrm{SNR}_\mathrm{R} |{\mathbm{f}}^H {{\mathbm{G}}_0} {\mathbm{x}}|^2}{\sum\limits_{q=1}^Q{ \mathrm{CNR}_{\mathrm{R},q} |{\mathbm{f}}^H   {{\mathbm{G}}_q} {\mathbm{x}}|^2 } + \left\| {\mathbm{f}} \right\|_F^2} ,
	\label{eq:23}
\end{equation}
with $\mathrm{SNR}_\mathrm{R} = \varsigma_0^2 / \sigma_\mathrm{R}^2 $ and $\mathrm{CNR}_{\mathrm{R},q} = 2\varsigma_q^2 / \pi \sigma_\mathrm{R}^2 $.

\subsubsection{Radar Sensing Metric}

Based on above derivation, we have the following two insights:

\begin{insight}[1BitDFRC Radar Metric]\label{ins:1}
	As indicated in \eqref{eq:22}, with $N_{\mathrm{R}} \!\!\to\!\! \infty$, the probability of detection $\mathrm{Pr}_\mathrm{D}$ for the considered 1BitDFRC system with 1-bit ADCs is a function of quantized signal-clutter-plus-noise ratio (QSCNR). 
	Specifically, when the false-alarm probability $\mathrm{Pr}_{\mathrm{FA}} \in (0 , 1)$ is fixed, we have $\ln(\mathrm{Pr}_{\mathrm{FA}})<0$ and thus $\mathrm{Pr}_\mathrm{D}$ monotonically increases with $\mathrm{QSCNR}$. 
	Therefore, maximizing $\text{QSCNR}({\mathbm{f}}, \mathbm{x})$, or ensuring that it remains no less than a predefined threshold, can improve or maintain radar performance.
\end{insight}

\begin{insight}[1BitDFRC Performance Loss Caused by 1-Bit ADC Quantization]\label{ins:2}
	The probability of detection $\mathrm{Pr}_\mathrm{D}$ for DFRC with $\infty$-bit ADCs and 1BitDFRC with 1-bit ADCs follows the same form \cite{cui2013mimo}. 
	Upon examining the signal-clutter-plus-noise ratio (SCNR) in \cite{cui2013mimo} and QSCNR in \eqref{eq:23}, they satisfy (for $N_{\mathrm{R}}\to\infty$)
	$ \mathrm{QSCNR} \approx \frac{2}{\pi}\,\mathrm{SCNR}$, 
	where the factor $2/\pi$ comes from 1-bit quantization.
	Therefore, in the dB scale, 1BitDFRC incurs an SCNR loss of
	$ 10\log_{10}(\mathrm{SCNR})-10\log_{10}(\mathrm{QSCNR}) \approx 10\log_{10}\!({\pi} / {2})\approx 1.96~\mathrm{dB}$, 
	due to 1-bit ADC quantization when $N_{\mathrm{R}} \to \infty$.
\end{insight}

\begin{remark}
	The state-of-the-art literature \cite{stein2013quantization,mo2017hybrid,roth2017achievable} in the communications domain has also concluded that 1-bit ADCs result in a 1.96dB performance loss.
	These works \cite{stein2013quantization,mo2017hybrid,roth2017achievable} are based on the conventional Bussgang model \cite{papoulis1965random}, which approximates the quantized signal after 1-bit ADCs using a linear model. 
	In contrast, our analysis is based on statistical reasoning and independently arrives at the same conclusion. 
	This highlights the fundamental distinction between our approach and existing work.
\end{remark}

\begin{remark}
		In above formulations and \textit{Insight \ref{ins:2}}, $N_{\mathrm R}\to\infty$ is introduced only to obtain a tractable asymptotic analysis. 
		In practice, $N_{\mathrm R}$ is large but finite, and our simulations (See Section \ref{Sec:6}) show that a moderate array size already well matches the asymptotic analysis, indicating that our formulations and \textit{Insight \ref{ins:2}} remain valid without requiring an excessively large $N_{\mathrm{R}}$. 
		This implies that employing 1-bit ADCs in realistic massive MIMO is theoretically and practically efficient. 
\end{remark}

\textit{Insight \ref{ins:1}} reveals that it is reasonable to choose the QSCNR as the radar performance metric for 1BitDFRC with massive MIMO.
Moreover, \textit{Insight \ref{ins:2}} indicates that the use of 1-bit ADCs results in only a $1.96$ dB SCNR degradation, yet achieves significant savings in hardware cost and power consumption for massive MIMO, thereby confirming the effectiveness of 1-bit ADC adoption.

\vspace{-1em}
\subsection{Communication BER Performance and Metric}\label{Sec:3-2}
This subsection motivates the communication metric used in 1BitDFRC.
We start from the well-known CI-based SLP for $\mathcal{M}$-PSK\footnote{This paper focus on $\mathcal{M}$-PSK. However, the SEP analysis and CI-based SLP design can be directly generalized to quadrature amplitude modulation (QAM) \cite{shao2019framework,li20211} as well.}, and then establish an explicit link between the CI-margin and the communication SEP.

\subsubsection{CI-Margin in CI-based SLP}
Consider the received model in \eqref{eq:8}. Following the CI-based SLP framework for $\mathcal{M}$-PSK signaling, we first rotate the received signal by the phase of the desired symbol and define $\tilde{y}_u \triangleq y_u e^{-\jmath\angle(s_u)}$.
Under nearest-neighbor detection, correct detection occurs when $\tilde{y}_u$ lies within the corresponding $\mathcal{M}$-PSK decision sector.
Accordingly, the detection reliability can be quantified by the \emph{CI-margin} (also referred to as the safe margin), i.e., the minimum distance to the two decision boundaries.

Specifically, define $\bar{\mathbm{h}}_u \triangleq \mathbm{h}_u e^{\jmath\angle(s_u)}$, $\kappa_1 = 1 - e^{-\jmath\frac{\pi}{2}}\cot (\frac{\pi}{\mathcal{M}} )$ and
$\kappa_2 = 1 + e^{-\jmath\frac{\pi}{2}}\cot (\frac{\pi}{\mathcal{M}} )$.
Then, the CI-margin is given by
\begin{equation}\label{eq:27-n}
	\alpha_u = \min \{ \Re ( \kappa_1 \bar{\mathbm{h}}_u^H \mathbm{x} ) , \Re ( \kappa_2 \bar{\mathbm{h}}_u^H \mathbm{x} )  \} .
\end{equation}

\begin{figure}[t]
	\centering
	\includegraphics[width=0.6\linewidth]{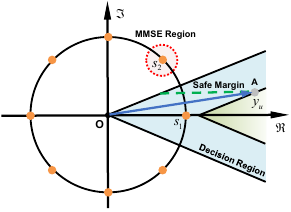}
	\vspace{-1em}
	\caption{Illustration of the decision region, CI-margin, and the MMSE region of a transmit symbol selected from the 8-PSK constellation.}
	\label{fig:SLB}
	\vspace{-1em}
\end{figure}

For intuition, an 8-PSK constellation example is illustrated in Fig.~\ref{fig:SLB}. 
The two green dotted lines indicate the distances from the noise-free received point to the two decision boundaries; the CI-margin corresponds to the smaller of these two distances.

\subsubsection{SEP Bound}
In addition to above intuition, we formally establish the following theorem, which bounds the SEP in terms of the CI-margin $\alpha_u$.
\begin{theorem}\label{the:2}
	For $\mathcal{M}$-PSK signaling under the receive model \eqref{eq:8}, the SEP is bounded as
	\begin{equation}\label{eq:26}
		\Psi \left( \frac{\sqrt{2}\,\sin(\frac{\pi}{\mathcal{M}})\,\alpha_u}{\sigma_{\mathrm{C},u}} \right)
		\le \mathrm{SEP}_u \le
		2 \Psi \left( \frac{\sqrt{2}\,\sin(\frac{\pi}{\mathcal{M}})\,\alpha_u}{\sigma_{\mathrm{C},u}} \right),
	\end{equation}
	where $\Psi(x) = \frac{1}{\sqrt{2\pi}} \int_x^\infty e^{-t^2/2}\mathrm{d}t$ and $\alpha_u$ is given in \eqref{eq:27-n}.
\end{theorem}
\begin{IEEEproof}
	Please refer to \cite[Sec. 8.1.1.3]{simon2004digital} and \cite{shao2019framework,shao2018multiuser,Wu2023Diversity}.
\end{IEEEproof}

Based on above theorem, we have following design insight.
\begin{insight}[Design Guideline]
	Since $\Psi(\cdot)$ is monotonically decreasing, the SEP can be improved by enlarging the CI-margin $\alpha_u$. 
	Therefore, the communication performance of 1BitDFRC can be guaranteed by maximizing $\alpha_u$.
	Alternatively, to explicitly control the communication QoS, we may enforce a CI-margin constraint $\alpha_u \ge \lambda_u$, where $\lambda_u$ is chosen according to a target SNR requirement as $\lambda_u \triangleq \sqrt{\Gamma_u \sigma_{\mathrm{C},u}^2}$ \cite{li2020interference,li2018massive,li20211,Wu2023CI}, with $\Gamma_u$ denoting the desired communication SNR.
\end{insight}

\subsubsection{Comparison to Conventional Metric}
In 1-bit communication scenarios, the MMSE is a widely used communication metric \cite{cheng2021transmit,yu2022precoding,castaneda20171}, aiming to ensure the received symbol $y_u$ is close to the desired symbol $s_u$ by mitigating  multiuser interference (MUI).
This restricts the received symbol $y_u$ within a circular region (red circular region in Fig. \ref{fig:SLB}) around $s_u$, which limits design flexibility. 
In contrast, the \textit{Theorem \ref{the:2}} indicates that as long as the received signal remains within the CI-margin, it can guarantee the communication BER performance, offering a new perspective to ensure communication BER performance.

Based on the previous analyses and guidelines, in the following two sections, we shall reformulate the QoS-constrained and QoD-constrained design problems, and present the proposed solutions, along with convergence analysis.

\section{QoS-Constrained 1BitDFRC Design}\label{Sec:4}
\subsection{Problem Formulation}
Building on the analysis and metrics in Sec.~\ref{Sec:3}, the QoS-constrained 1BitDFRC design problem in \eqref{eq:QoS} can be reformulated as
\begin{subequations}
	\begin{align}
		\mathop {\max }\limits_{{\mathbm{x}},{{\mathbm{f}}}} \quad & \mu \frac{  {\mathbm{f}}^H {\mathbm G}_0 {\mathbm x} {\mathbm x}^H {\mathbm G}_0^H {\mathbm{f}}  }{  \sum_{q=1}^{Q}{ \mathrm{CNR}_{\mathrm{R},q} {\mathbm{f}}^H {\mathbm G}_q {\mathbm x} {\mathbm x}^H {\mathbm G}_q^H {\mathbm{f}} } + \| {\mathbm{f}} \|_F^2 } \label{eq:28a} \\
		{\text{s}}.{\text{t}}. \quad & \min \{ \Re ( \kappa_1 \bar{\mathbm{h}}_u^H \mathbm{x} ) , \Re ( \kappa_2 \bar{\mathbm{h}}_u^H \mathbm{x} )  \} \ge \lambda_u , \forall u , \label{eq:28b} \\
		& {\mathbm{x}} \in \mathcal{X}_{\text{DAC}}^{\text{1-Bit}} = \left\{ \frac{E}{{\sqrt {2{M_{\text{T}}}} }}  ({ \pm 1 \pm \jmath}) \right\} , \label{eq:28c}
	\end{align}
	\label{eq:28}%
\end{subequations}
where $\mu = 2\mathrm{SNR}_\mathrm{R}/\pi$ is a constant.

Before detailing the solutions to problem \eqref{eq:28}, some transformations will be performed.  
It is clear that problem \eqref{eq:28} involves two variables, $\mathbm{x}$ and ${\mathbm{f}}$. 
Additionally, we observe that the receive filter ${\mathbm{f}}$ is only present in the objective of problem \eqref{eq:28}. 
Therefore, updating the receive filter ${\mathbm{f}}$ leads to the following unconstrained problem
\begin{equation}
	\mathop {\max }\limits_{{{\mathbm{f}}}} \quad  \mu \frac{  {\mathbm{f}}^H {\mathbm G}_0 {\mathbm x} {\mathbm x}^H {\mathbm G}_0^H {\mathbm{f}}  }{  \sum_{q=1}^{Q}{ \mathrm{CNR}_{\mathrm{R},q} {\mathbm{f}}^H {\mathbm G}_q {\mathbm x} {\mathbm x}^H {\mathbm G}_q^H {\mathbm{f}} } + \| {\mathbm{f}} \|_F^2 } ,
\end{equation}
which is a typical Rayleigh quotient problem with closed-form solutions as
\begin{equation}\label{eq:31}
	{{\mathbm{f}}} = \frac{ {{\Big[ {\sum\limits_{q = 1}^Q { \mathrm{CNR}_{\mathrm{R},q} {{\mathbm{G}}_q}{\mathbm x}{\mathbm{x}}^H{\mathbm{G}}_q^H}  + {\mathbm{I}_{N_{\mathrm{R}}}}} \Big]^{ - 1}}{{\mathbm{G}}_0}{\mathbm{x}}} } { {{{\mathbm{x}}^H}{\mathbm{G}}_0^H{\Big[ {\sum\limits_{q = 1}^Q { \mathrm{CNR}_{\mathrm{R},q} {{\mathbm{G}}_q}{\mathbm x}{\mathbm{x}}^H{\mathbm{G}}_q^H}  + {\mathbm{I}_{N_{\mathrm{R}}}}} \Big]^{ - 1}}{{\mathbm{G}}_0}{\mathbm{x}}} } .
\end{equation}

Substituting ${\mathbm{f}}$ from \eqref{eq:31} into the objective function of \eqref{eq:28}, the optimization problem in \eqref{eq:28} is equivalently expressed as
\begin{subequations}
	\begin{align}
		\mathop {\max }\limits_{{\mathbm{x}}} \;\; & \text{QSCNR}({\mathbm x}) \label{eq:32a} \\
		{\text{s}}.{\text{t}}. \;\; & \min \{ \Re ( \kappa_1 \bar{\mathbm{h}}_u^H \mathbm{x} ) , \Re ( \kappa_2 \bar{\mathbm{h}}_u^H \mathbm{x} )  \} \ge \lambda_u , \forall u , \label{eq:32b} \\
		& {\mathbm{x}} \in \mathcal{X}_{\text{DAC}}^{\text{1-Bit}} = \left\{ \frac{E}{{\sqrt {2{N_{\text{T}}}} }}  ({ \pm 1 \pm \jmath}) \right\},\label{eq:32c}
	\end{align}
	\label{eq:32}%
\end{subequations}
where $\text{QSCNR}({\mathbm x})$ is given by
\begin{equation}
	\begin{aligned}
		& \text{QSCNR}({\mathbm x}) \\
		& = \mu {{{\mathbm{x}}^H}{\mathbm{G}}_0^H{\Big[ {\sum\limits_{q = 1}^Q { \mathrm{CNR}_{\mathrm{R},q} {{\mathbm{G}}_q}{\mathbm x}{\mathbm{x}}^H{\mathbm{G}}_q^H}  + {\mathbm{I}_{N_{\mathrm{R}}}}} \Big]^{ - 1}}{{\mathbm{G}}_0}{\mathbm{x}}} .
	\end{aligned}
\end{equation}

Now, we end up with an optimization problem involving a single variable, i.e., $\mathbm{x}$.
The remaining challenges in solving \eqref{eq:32} lie in the complicated objective function and the discrete values of $\mathbm{x}$.
In the next subsection, we will address these challenges and present the solution to problem \eqref{eq:32}.

\vspace{-0.5em} \subsection{Proposed Solution to Problem \eqref{eq:32}}\label{Sec_4_B}

\subsubsection{MM for Objective Function \eqref{eq:32a}}
The objective function \eqref{eq:32a} is a complex fractional function with inverse operation. 
We apply MM approaches to simplify \eqref{eq:32a} into a more tractable linear function. 
Specifically, we first present the following proposition to address the inverse operation in \eqref{eq:32}.
\begin{proposition}\label{pro:1}
	Let ${\mathbm M}$ be a Hermitian matrix, a minorizer of $\mathbm{s}^H\mathbm{M}^{-1}\mathbm{s}$ is given by
	\begin{equation}
		\mathbm{s}^H\mathbm{M}^{-1}\mathbm{s} \ge  2  \Re \{\mathbm{s}_t^H\mathbm{M}_t^{-1}\mathbm{s} \}  - \text{Tr}\left\{ \mathbm{M}_t^{-1} \mathbm{s}_t \mathbm{s}_t^H \mathbm{M}_t^{-1}\mathbm{M} \right\} ,
		\label{eq:33}
	\end{equation}
	where $( \mathbm{s}_t ,  \mathbm{M}_t )$ represent the point at $t$-th iteration.
\end{proposition}
\begin{IEEEproof}
	See \cite[Section III]{sun2016majorization}.
\end{IEEEproof}

By submitting $\mathbm{M} = {\sum_{q = 1}^Q { \mathrm{CNR}_{\mathrm{R},q} {{\mathbm{G}}_q}{\mathbm x}{\mathbm{x}}^H{\mathbm{G}}_q^H}  + {\mathbm{I}_{N_{\mathrm{R}}}}} $ and $\mathbm{s} = \mathbm{G}_0 \mathbm{x}$ into \textit{Proposition \ref{pro:1}}, we have 
\begin{subequations}\label{eq:34}
	\begin{align}
		& \text{QSCNR}({\mathbm x}) \ge   - \mu \text{Tr}\left\{ \mathbm{M}_t^{-1} \mathbm{G}_0 \mathbm{x}_t \mathbm{x}_t^H \mathbm{G}_0^H \mathbm{M}_t^{-1} \mathbm{M} \right\}   \nonumber \\
		& \quad +  2\mu \Re \left\{ \mathbm{x}_t^H\mathbm{G}_0^H\mathbm{M}_t^{-1}\mathbm{G}_0\mathbm{x} \right\} + \text{const}_1 \\
		& = - \mu \mathbm{x}^H \widetilde{\mathbm{M}}_t \mathbm{x} + 2 \mu \Re \left\{ \mathbm{x}_t^H\mathbm{G}_0^H\mathbm{M}_t^{-1}\mathbm{G}_0\mathbm{x} \right\} + \text{const}_1 \\
		& = \widehat{\text{QSCNR}}({\mathbm x} | {\mathbm{x}}_t) ,
	\end{align}
\end{subequations}
where $\text{const}_1 = - \mu \text{Tr} \{ \mathbm{M}_t^{-1} \mathbm{G}_0 \mathbm{x}_t \mathbm{x}_t^H \mathbm{G}_0^H \mathbm{M}_t^{-1} \}$
and  $\widetilde{\mathbm{M}}_t = \sum_{q=1}^Q \mathrm{CNR}_{\mathrm{R},q} {\mathbm{G}}_q^H \mathbm{M}_t^{-1} \mathbm{G}_0 \mathbm{x}_t \mathbm{x}_t^H \mathbm{G}_0^H \mathbm{M}_t^{-1} {\mathbm{G}}_q $.

So far, we have transformed the complicated objective function in \eqref{eq:32a} into a quadratic function $\widehat{\text{QSCNR}}({\mathbm x} | {\mathbm{x}}_t)$ by using MM approach.
To further simplify $\widehat{\text{QSCNR}}({\mathbm x} | {\mathbm{x}}_t)$, we can apply the MM approach again and have the following proposition.
\begin{proposition}\label{S3_pro:3}
	Let ${\mathbm J}$ be a Hermitian matrix, a majorizer of $\mathbm{x}^H\mathbm{J}\mathbm{x}$ can be given by
	\begin{equation}
		\begin{aligned}
			\mathbm{x}^H\!\mathbm{J}\mathbm{x} \le & \varpi_{\text{m}}  \left\| \mathbm{x} \right\|_F^2  + 2\Re\!\left\{ \mathbm{x}_t^H ( \mathbm{J}  - \varpi_{\text{m}} \mathbm{I} ) \mathbm{x} \right\} \\
			 &  - \mathbm{x}_t^H ( \mathbm{J}-  \varpi_{\text{m}} \mathbm{I} ) \mathbm{x}_t ,
		\end{aligned}
	\end{equation}
	where $\varpi_{\text{m}}$ represents the maximum eigenvalue of $\mathbm{J}$, and $\mathbm{I}$ denotes the identity matrix of appropriate dimension.
\end{proposition}
\begin{IEEEproof}
	See \cite[Section III-C, Example 13]{sun2016majorization}.
\end{IEEEproof}
Therefore, by defining $\mathbm{J} = \widetilde{\mathbm{M}}_t$ and recognizing $\|\mathbm{x}\|_F^2 = E$, we have
\begin{subequations}\label{eq:37}
	\begin{align}
		& \text{QSCNR}({\mathbm x}) \ge \widehat{\text{QSCNR}} ({\mathbm x} | {\mathbm{x}}_t) \nonumber \\
		& \ge   -  2 \mu \Re\left\{ \mathbm{x}_t^H ( \widetilde{\mathbm{M}}_t - \ell_{\text{max}} \mathbm{I}_{N_{\mathrm{T}}} ) \mathbm{x} \right\} \\
		& \quad +  2 \mu \Re \left\{ \mathbm{x}_t^H\mathbm{G}_0^H\mathbm{M}_t^{-1}\mathbm{G}_0\mathbm{x} \right\}  \nonumber \\
		& \quad  - \mu \ell_{\text{max}}E  + \mu \mathbm{x}_t ( \widetilde{\mathbm{M}}_t - \ell_{\text{max}} \mathbm{I}_{N_{\mathrm{T}}} ) \mathbm{x}_t + \text{const}_1 \\
		& = \Re\left\{ \mathbm{w}_t^H \mathbm{x} \right\} + \text{const}_2 = \overline{ \text{QSCNR} } ({\mathbm x} | {\mathbm{x}}_t) ,
	\end{align}
\end{subequations}
where $\ell_{\text{max}}$ is maximum eigenvalue of $\widetilde{\mathbm{M}}_t$, $\mathbm{w}_t$ is defined as $\mathbm{w}_t = 2 \mu \mathbm{G}_0 \mathbm{M}_t^{-1} \mathbm{G}_0^H \mathbm{x}_t - 2 \mu ( \widetilde{\mathbm{M}}_t - \ell_{\text{max}} \mathbm{I}_{N_{\mathrm{T}}} ) \mathbm{x}_t $
and %$\text{const}_2$ is constant with definition as 
	$\text{const}_2 = \text{const}_1 - \mu \ell_{\text{max}}E + \mu \mathbm{x}_t ( \widetilde{\mathbm{M}}_t - \ell_{\text{max}} \mathbm{I}_{N_{\mathrm{T}}} ) \mathbm{x}_t $.

By the transformation from \eqref{eq:34} to \eqref{eq:37}, we have simplified $\text{QSCNR}({\mathbm x})$ to a linear function $\overline{ \text{QSCNR} } ({\mathbm x} | {\mathbm{x}}_t)$, significantly reducing the design complexity.

\subsubsection{Rearrangement for QoS Constraint \eqref{eq:32b}}
We propose to rearrange \eqref{eq:32b} to facilitate the design. 
To this end, by noting $\min\{a , b\} \ge c \Rightarrow \{a \ge c , b \ge c\}$, we have
\begin{equation}\label{NEQ_29_new}
	\begin{aligned}
		\eqref{eq:32b} & \Leftrightarrow \min \{ \Re ( \kappa_1 \bar{\mathbm{h}}_u^H \mathbm{x} ) , \Re ( \kappa_2 \bar{\mathbm{h}}_u^H \mathbm{x} )  \} \ge \lambda_u  , \\
		& \Leftrightarrow \left\{
			\Re \left\{ \kappa_1 \bar{\mathbm{h}}_u^H \mathbm{x} \right\} \ge \lambda_u  , 
			\Re \left\{ \kappa_2 \bar{\mathbm{h}}_u^H \mathbm{x} \right\} \ge \lambda_u 
		\right\}  ,  \\ 
		& \Leftrightarrow \Re \left\{ \mathbm{A}_u \mathbm{x} \right\} \ge \lambda_u  \mathbm{1}_2 = \mathbm{\lambda}_u ,
	\end{aligned}
\end{equation}
where
$\mathbm{1}_N \in \mathbb{R}^N$ is a summing vector with all entries 1, $\mathbm{A}_u = [\kappa_1 , \kappa_2 ]^T \otimes \bar{\mathbm{h}}_u^H$, and $\mathbm{\lambda}_u \nonumber = \lambda_u \mathbm{1}_2$.

\subsubsection{Integer Linear Programming (ILP) for \eqref{eq:32}}\label{Sec_4_B_3}
Based on the above reformulations, the original problem \eqref{eq:32} can be transformed into
\begin{subequations}
	\begin{align}
		\mathop {\max }\limits_{{\mathbm{x}} } \quad &  \Re\left\{ \mathbm{w}_t^H \mathbm{x} \right\}  \label{eq:41a} \\
		{\text{s}}.{\text{t}}. \quad & \Re \left\{ \mathbm{A} \mathbm{x} \right\} \ge  \mathbm{\lambda} , \label{eq:41b}  \\
		& {\mathbm{x}} \in \mathcal{X}_{\text{DAC}}^{\text{1-Bit}} = \left\{ \frac{E}{{\sqrt {2{N_{\text{T}}}} }}  ({ \pm 1 \pm \jmath}) \right\} , \label{eq:41c} 
	\end{align}
	\label{eq:41}%
\end{subequations}
where $\mathbm{A} = [\mathbm{A}_1^T , \cdots , \mathbm{A}_U^T ]^T$ and $\mathbm{\lambda} = [\mathbm{\lambda}_1^T , \cdots , \mathbm{\lambda}_U^T]^T$.
Now, the challenge in solving \eqref{eq:41} lies in the discrete constraint \eqref{eq:41c}.
To address this, we reformulate problem \eqref{eq:41} into its real form as follows
\begin{subequations}
	\begin{align}
		\mathop {\max }\limits_{\widetilde{{\mathbm{x}}} } \quad &  \Re\left\{ \widetilde{\mathbm{w}}_t^T \widetilde{\mathbm{x}} \right\}  \label{eq:43a} \\
		{\text{s}}.{\text{t}}. \quad\;\; &  \widetilde{\mathbm{A}} \widetilde{\mathbm{x}}  \ge  \widetilde{\mathbm{\lambda}} , \label{eq:43b} \\
		& \widetilde{{\mathbm{x}}} \in \mathcal{X}_{\text{DAC},\Re}^{\text{1-Bit}} =  \left\{ - \frac{E}{{\sqrt {2{N_{\text{T}}}} }}  , + \frac{E}{{\sqrt {2{N_{\text{T}}}} }} \right\} . \label{eq:43c}
	\end{align}
	\label{eq:43}%
\end{subequations}
where $\widetilde{\mathbm{x}} = [\Re\{ \mathbm{x}^T \} , \Im\{ \mathbm{x}^T \}]^T$, $\widetilde{\mathbm{w}} = [\Re\{ \mathbm{w}^T \} , - \Im\{ \mathbm{w}^T \}]^T$ and $\widetilde{\mathbm{A}} = \left[ \begin{array}{ll}
		\Re\{ \mathbm{A} \} & -\Im\{ \mathbm{A} \} \\
		\Im\{ \mathbm{A} \} & \Re\{ \mathbm{A} \}
\end{array}\right]$.
Problem \eqref{eq:43} is a standard ILP with binary constraint \eqref{eq:43c}, which can be optimally solved using many existing solvers, such as branch and bound (BnB), implicit enumeration methods, and others.
Here, we adopt the BnB method to tackle \eqref{eq:43}. 
For details of the BnB method, interested readers are referred to \cite{li2020interference,landau2017branch,boyd2007branch}.

Finally, we present the pseudo code of the proposed solution to QoS-constrained 1BitDFRC design in Algorithm \ref{alg:1}.

\begin{algorithm}[t]
	\SetAlgoLined
	\SetKwInOut{Input}{Input}\SetKwInOut{Output}{Output}
	\Input{System Parameter}
	\Output{${\mathbm{f}}$ and $\mathbm{x}$}
	\textbf{Initialization:} Set $\mathbm{x}_0$ and $t = 0$\;
	\While{No Convengence}{
		$t = t + 1$\;
		Update ${\mathbm{x}}_{t}$ by solving \eqref{eq:43} \;
	}
	${\mathbm{x}} = {\mathbm{x}}_{t}$\;
	${{\mathbm{f}}} = \frac{ {{{\left[ {\sum_{q = 1}^Q { \mathrm{CNR}_{\mathrm{R},q} {{\mathbm{G}}_q}{\mathbm x}{\mathbm{x}}^H{\mathbm{G}}_q^H}  + {\mathbm{I}_{N_{\mathrm{R}}}}} \right]}^{ - 1}}{{\mathbm{G}}_0}{\mathbm{x}}} } { {{{\mathbm{x}}^H}{\mathbm{G}}_0^H{{\left[ {\sum_{q = 1}^Q { \mathrm{CNR}_{\mathrm{R},q} {{\mathbm{G}}_q}{\mathbm x}{\mathbm{x}}^H{\mathbm{G}}_q^H}  + {\mathbm{I}_{N_{\mathrm{R}}}}} \right]}^{ - 1}}{{\mathbm{G}}_0}{\mathbm{x}}} }$ .
	\caption{Proposed solution to QoS-constrained 1BitDFRC design problem \eqref{eq:28}.}
	\label{alg:1}
\end{algorithm}

\vspace{-0.5em} \subsection{Summary}

The convergence of the proposed Algorithm~\ref{alg:1} is guaranteed by the following theorem.
\begin{theorem}\label{the:3}
	Let $\{\text{QSCNR}(\mathbm{x}_t)\}$ denote the sequence of objective values generated by Algorithm.~\ref{alg:1}. This sequence is non-decreasing and converges to a local maximum.
\end{theorem}
\begin{IEEEproof}
	See Appendix \ref{proof:3}.
\end{IEEEproof}

Finally, we note although the worst-case complexity of the BnB-based algorithm is exponential, the use of efficient pruning mechanisms significantly reduces the practical search space, making it computationally tractable for massive MIMO settings (e.g., $N_{\mathrm{T}} \le 512$) \cite{das2024branch}. 
It is worth noting that even conventional convex methods face scalability bottlenecks in ultra-large-scale antenna systems (e.g., $N_{\mathrm{T}} \ge 1000$) due to $\mathcal{O}(N^3)$ matrix operations, primarily stemming from matrix inversions and SVD. 
Consequently, the development of fast and efficient design methods for such systems remains an open problem.

\section{QoD-Constrained 1BitDFRC Design}\label{Sec:5}

In this section, the method proposed in Sec.~\ref{Sec:4} is extended to address the QoD-constrained 1BitDFRC design in \eqref{eq:QoD}.

\vspace{-0.5em} \subsection{Problem Reformulation}
Similarly, guided by the performance analysis in Sec. \ref{Sec:3}, the QoD-constrained 1BitDFRC design in \eqref{eq:QoD} can be recast as
\begin{subequations}
	\begin{align}
		\mathop {\max }\limits_{{\mathbm{x}},{{\mathbm{f}}}}  &  \; \min_u \{ \Re ( \kappa_1 \bar{\mathbm{h}}_u^H \mathbm{x} ) , \Re ( \kappa_2 \bar{\mathbm{h}}_u^H \mathbm{x} )  \}  \label{eq:44a} \\
		{\text{s}}.{\text{t}}. \; & \frac{2}{\pi} \frac{ \mathrm{SNR}_\mathrm{R} {\mathbm{f}}^H {\mathbm G}_0 {\mathbm x} {\mathbm x}^H {\mathbm G}_0^H {\mathbm{f}}  }{  \sum_{q=1}^{Q}{ \mathrm{CNR}_{\mathrm{R},q} {\mathbm{f}}^H {\mathbm G}_q {\mathbm x} {\mathbm x}^H {\mathbm G}_q^H {\mathbm{f}} } + \| {\mathbm{f}} \|_F^2 } \!\ge\! \chi , \label{eq:44b} \\
		& {\mathbm{x}} \in \mathcal{X}_{\text{DAC}}^{\text{1-Bit}} = \left\{ \frac{E}{{\sqrt {2{M_{\text{T}}}} }}  ({ \pm 1 \pm \jmath}) \right\} , \label{eq:44c}
	\end{align}
	\label{eq:44}%
\end{subequations}
where $\chi \ge 0$ represents the radar QSCNR threshold.
For the desired $\ln \mathrm{Pr}_\mathrm{FA}$ and $\mathrm{Pr}_\mathrm{D}$,  $\chi \ge 0$ can be determined by $\chi = \ln \mathrm{Pr}_\mathrm{FA} / \mathrm{Pr}_\mathrm{D} - 1$.

\vspace{-0.5em} \subsection{Proposed Solution to \eqref{eq:44}}
Similarly to problem \eqref{eq:28}, after introducing auxiliary variable $\lambda$, applying the MM to the constraint set \eqref{eq:44b} and rearranging \eqref{eq:44a}, problem \eqref{eq:44} can be transformed into a simpler form as follows
\begin{subequations}
	\begin{align}
		\mathop {\max }\limits_{\mathbm{x} , \lambda} \quad &  \lambda  \label{eq:45a} \\
		{\text{s}}.{\text{t}}. \quad & \Re \left\{ \mathbm{A} \mathbm{x} \right\} \ge  \lambda \mathbm{1}_{2U} , \label{eq:45b} \\
		& \overline{ \text{QSCNR} } ({\mathbm x} | {\mathbm{x}}_t)  = \Re\left\{ \mathbm{w}_t^H \mathbm{x} \right\} + \text{const}_2 \ge \chi , \label{eq:45c} \\
		& {\mathbm{x}} \in \mathcal{X}_{\text{DAC}}^{\text{1-Bit}} = \left\{ \frac{E}{{\sqrt {2{N_{\text{T}}}} }}  ({ \pm 1 \pm \jmath}) \right\} , \label{eq:45d}
	\end{align}
	\label{eq:45}%
\end{subequations}
where $\mathbm{w}_t$, $\mathbm{A}$, and $\text{const}_2$ share the same formulas as those in Sec. \ref{Sec:4}.
By converting problem \eqref{eq:45} into its real-valued form, we have the following standard ILP problem
\begin{subequations}
	\begin{align}
		\mathop {\max }\limits_{\widetilde{\mathbm{x}} , \lambda } \quad &  \lambda  \label{eq:46a} \\
		{\text{s}}.{\text{t}}. \quad &  \widetilde{\mathbm{A}} \widetilde{\mathbm{x}}  \ge \lambda {\mathbm{1}}_{2U} , \label{eq:46b} \\
		& \Re\left\{ \widetilde{\mathbm{w}}_t^T \widetilde{\mathbm{x}} \right\} + \text{const}_2 \ge \chi , \label{eq:46c} \\
		& \widetilde{{\mathbm{x}}} \in \left\{ - \frac{E}{{\sqrt {2{N_{\text{T}}}} }}  , + \frac{E}{{\sqrt {2{N_{\text{T}}}} }} \right\} , \label{eq:46d}
	\end{align}
	\label{eq:46}%
\end{subequations}
which can also be optimally solved by BnB solver.
Since the algorithm workflow and convergence analysis to problem \eqref{eq:44} follows the same procedure as described in Sec. \ref{Sec:4}, the details are omitted for simplicity.

\section{Further Extension}
The radar sensing performance analysis and joint designs presented above are based on the single-target model in \eqref{eq_3_New}. 
In this section, we extend the analysis of radar sensing performance under 1-bit DACs, along with the joint transceiver design, to multi-target scenarios.

\subsection{Radar Model and Radar Sensing Performance}
Consider a set of $K$ targets of interest located at angles $\theta_{\mathrm{T}, k}, \forall k$ in the presence of $Q$ stationary clutter sources at angles $\theta_{\mathrm{C}, q}, \forall q$. 
Accordingly, the overall observation at the DFBS can be expressed as
\begin{equation}
	\begin{aligned}
		\mathbm{r} = & \mathbm{r}_\mathrm{T} + \mathbm{r}_\mathrm{C} + \mathbm{n}_\mathrm{R} = \sum_{k=1}^{K} \underbrace { \gamma_{\mathrm{T}, k}\mathbm{G}_{\mathrm{T}, k} {\mathbm{x}} }_{\mathbm{r}_{\mathrm{T}, k}}
		+ \sum\limits_{q = 1}^Q \underbrace {{{\gamma_{\mathrm{C},q}}\mathbm{G}_{\mathrm{C},q}{\mathbm{x}}} }_{ \mathbm{r}_{\mathrm{C},q} } + \mathbm{n}_\mathrm{R} ,
	\end{aligned}
\end{equation}
where the definitions of $\gamma_{\mathrm{T}, k}$, $\gamma_{\mathrm{C}, q}$, $\mathbm{G}_{\mathrm{T}, k}$, and $\mathbm{G}_{\mathrm{C}, q}$ are consistent with those provided in Sec. \ref{Sec:2_B}.

Then, the received signal $\mathbm{r}$ is quantized by 1-bit ADCs, i.e., $\widetilde{\mathbm{r}} = \mathcal{Q}_{\text{ADC}}^{\text{1-Bit}} ( \mathbm{r} )$, and processed by $K$ receive filters $\mathbm{f}_k$, $k = 1, \ldots, K$, with one filter dedicated to each target.
Therefore, the binary hypothesis test for target-$k$ detection is formulated as
\begin{equation}
	\left\{  
	\begin{array}{ll}
		\mathcal{H}_0^k: & z_k = {\mathbm{f}}_k^H\widetilde{\mathbm{r}} = {\mathbm{f}}_k^H \mathcal{Q}_{\text{ADC}}^{\text{1-Bit}}( \mathbm{r}_\mathrm{C} + \mathbm{n}_\mathrm{R} ) \\
		\mathcal{H}_1^k: & z_k = {\mathbm{f}}_k^H\widetilde{\mathbm{r}} = {\mathbm{f}}_k^H \mathcal{Q}_{\text{ADC}}^{\text{1-Bit}}( \mathbm{r}_\mathrm{T} + \mathbm{r}_\mathrm{C} + \mathbm{n}_\mathrm{R} )
	\end{array}
	\right. .
\end{equation}

Following the binary hypothesis test established above and the formulation procedures outlined in Sec. \ref{Sec:3-1}, th probability of detection $\mathrm{Pr}_{\mathrm{D}}^k$ of the $k$-th target can be evaluated as
\begin{equation}\label{eq_40_new}
	\begin{aligned}
		\mathrm{Pr}_\mathrm{D}^k &  = \exp\Bigg( \frac{\text{ln} (\mathrm{Pr}_\mathrm{FA})}{ 1 + \text{QSCNR}_k ( \{\mathbm{f}_k\} , \mathbm{x}) } \Bigg) ,
	\end{aligned}
\end{equation}
where $\text{QSCNR}_k(\{\mathbm{f}_k\} , \mathbm{x})$ is given in \eqref{eq_41_New}, as shown at the top of the next page.

\begin{figure*}
	\begin{equation}\label{eq_41_New}
		\begin{aligned}
			& \text{QSCNR}_k \left(\{\mathbm{f}_k\} , \mathbm{x}\right) = \frac{2}{\pi} \frac{\mathrm{SNR}_{\mathrm{R},k} |{\mathbm{f}}_k^H {{\mathbm{G}}_{\mathrm{T}, k}} {\mathbm{x}}|^2}{
				\sum\limits_{p=1,p\ne k}^K{ \mathrm{SNR}_{\mathrm{R},p} |\mathbm{f}_k^H   {{\mathbm{G}}_{\mathrm{T}, p}} {\mathbm{x}}|^2 }
				+ \sum\limits_{q=1}^Q{ \mathrm{CNR}_{\mathrm{R},q} |\mathbm{f}_k^H   {{\mathbm{G}}_{\mathrm{C}, q}} {\mathbm{x}}|^2 } 
				+ \left\| \mathbm{f}_k \right\|_F^2
			} .
		\end{aligned}
	\end{equation}
	\hrulefill
\end{figure*}

\subsection{Fairness-Aware QoS-Constrained 1BitDFRC Design}
Building on the above analysis and metrics, the fairness-aware QoS-constrained 1-bit DFRC design problem for the multi-target scenario can be reformulated as
\begin{subequations}\label{eq_42_new}
	\begin{align}
		\mathop {\max }\limits_{{\mathbm{x}},\{\mathbm{f}_k\}} \quad & \min_{k} \text{QSCNR}_k \left(\{\mathbm{f}_k\} , \mathbm{x}\right) \\
		{\text{s}}.{\text{t}}. \quad & \min \{ \Re ( \kappa_1 \bar{\mathbm{h}}_u^H \mathbm{x} ) , \Re ( \kappa_2 \bar{\mathbm{h}}_u^H \mathbm{x} )  \} \ge \lambda_u , \forall u , \\
		& {\mathbm{x}} \in \mathcal{X}_{\text{DAC}}^{\text{1-Bit}} = \left\{ \frac{E}{{\sqrt {2{M_{\text{T}}}} }}  ({ \pm 1 \pm \jmath}) \right\} .
	\end{align}
\end{subequations}
Problem \eqref{eq_42_new} guarantees that the output for all targets can be maximized, ensuring fairness among them. 
Note that \eqref{eq_42_new} shares the same structure as problem \eqref{eq:28}, and thus can be solved using a similar approach, the details of which are deferred to Appendix \ref{New_App_4}.

In addition to problem \eqref{eq_42_new}, one can also formulate a weighted sum-SCNR maximization problem to achieve fairness. 
As an extension here, we focus only on the max-min formulation \eqref{eq_42_new} as a representative example.
Besides, the QoD-constrained 1-bit DFRC design for the multi-target scenario can be formulated and solved similarly to the method shown in Sec. \ref{Sec:5}, which is also omitted here.

\section{Numerical Examples}\label{Sec:6}

In this section, numerical examples are provided to evaluate the performance of the 1BitDFRC.

\vspace{-0.5em} \subsection{System Setup}
Before presenting the simulation results, the parameter settings, benchmark approaches, and evaluation metrics are introduced in this subsection.

\subsubsection{Parameter Settings}
%Unless otherwise specified, in all simulations, the parameters are set as follows.
Consider a DFRC BS equipped with ${M_{\rm T}} = 128$ transmit and ${M_{\rm R}} = 128$ receive antennas, serving $U = 4$ downlink users. 
The available transmit power is $E = 1$ W, and identical noise power is assumed at all users, i.e., $\sigma_{\mathrm{C},u}^2 = \sigma_{\mathrm{C}}^2, \forall u$. 
For sensing, the DFRC BS attempts to detect a target located at $\theta_0 = 10^\circ$ in the presence of $Q = 2$ clutter sources situated at $\theta_1 = -50^\circ$ and $\theta_2 = 30^\circ$.
The radar SNR and CNR is set as $\mathrm{SNR}_\mathrm{R} = 15 \text{dB}$ and $\mathrm{CNR}_{\mathrm{R},q} = 30 \text{dB},\forall q$, respectively.

\subsubsection{Benchmark Approaches}
For comparison purposes, we include DFRC systems with different configurations as below:

\textbf{\textit{i):} DFRC with $\infty$-Bit DAC, $\infty$-Bit ADC. }
This configuration is known as the conventional fully-digital architecture. 
This benchmark can be obtained by modifying the proposed solutions in the following ways: replace the QSCNR with SCNR in \cite{cui2013mimo} and replace $\mathbm{x} \in \mathcal{X}_{\text{DAC}}^{\text{1-Bit}}$ with $\|\mathbm{x}\|_F^2 = E$.

\textbf{\textit{ii):} DFRC with $\infty$-Bit DAC, 1-Bit ADC.}
This benchmark can be obtained by replacing $\mathbm{x} \in \mathcal{X}_{\text{DAC}}^{\text{1-Bit}}$ with $\|\mathbm{x}\|_F^2 = E$ in the proposed solutions.

\textbf{\textit{iii):} DFRC with 1-Bit DAC, $\infty$-Bit ADC.}
This benchmark can be obtained by replacing the QSCNR with SCNR in \cite{cui2013mimo} in the proposed solutions.

For consistency in notation, we refer to the proposed 1BitDFRC system as the \textbf{DFRC with 1-Bit DAC and 1-Bit ADC}.

\subsubsection{Evaluation Metrics}
To thoroughly evaluate the performance of DFRC systems with different configurations, we define three evaluation metrics as follows:

\textbf{\textit{i):} Radar Energy Efficiency.}
To compare the energy efficiency (EE) for DFRC with different configurations, we define the radar EE as below
\begin{equation}
	\text{REE} = {\text{SCNR}} / {E_\text{tol}} ,
\end{equation}
where $E_\text{tol} = (N_\mathrm{T} + N_\mathrm{R})(E_\text{RF} + E_\text{LNA})  + 2E_\text{BB} + 2N_\mathrm{T}E_\text{ADC} + 2N_\mathrm{R}E_\text{DAC}$ is the total power consumption.
$E_\text{RF}$, $E_\text{LNA}$, and $E_\text{BB}$ represent the power consumption of the RF chain, low-noise amplifier, and baseband signal processing, respectively. 
These are set at 40 mW, 20 mW, and 200 mW.
The power consumption of ADCs and DACs follows the model in \cite{walden1999analog,mo2017hybrid}, where 10-bit resolution ($B=10$) is used to approximate the $\infty$-bit ADC/DAC case.

\textbf{\textit{ii):} Monte Carlo QSCNR.}
To validate the theoretical performance analysis (TA) presented in Sec. \ref{Sec:3-1}, we employ Monte Carlo (MC) simulations to calculate the practical QSCNR. 
Specifically, the theoretical QSCNR (TA) is calculated using \eqref{eq:23}, while the practical QSCNR (MC) is determined through $N_\mathrm{R}^\text{MC} = 10^6$ MC simulations.

\textbf{\textit{iii):} Monte Carlo BER.}
Similarly, to validate the performance TA in Sec. \ref{Sec:3-2}, the practical BER is calculated by MC with $N_\mathrm{C}^\text{MC} = 10^6$ being the number of MC simulations.

\vspace{-0.5em} \subsection{QoS-Constrained 1BitDFRC Design}
In this subsection, we examine the performance of the 1BitDFRC design proposed in Sec. \ref{Sec:4}, where the communication QoS requirement is constrained to facilitate comparison and validate radar performance.
We assume the 8-PSK modulated symbols are used in the simulations, i.e., $\mathcal{M}=8$.
We set the $\text{SNR}_{\mathrm{C},u} = {{E}}/{\sigma_{\mathrm{C},u}^2} = \text{5dB}, \forall u$.
The communication QoS threshold $\lambda_u$ are set the same, i.e., $\lambda_u = \sqrt{\sigma_c^2 \Gamma}, \forall u$ \cite{li2020interference}, with $\Gamma$ being the communication QoS threshold.

\begin{figure}[t]
	\centering
	\includegraphics[width=0.7\linewidth]{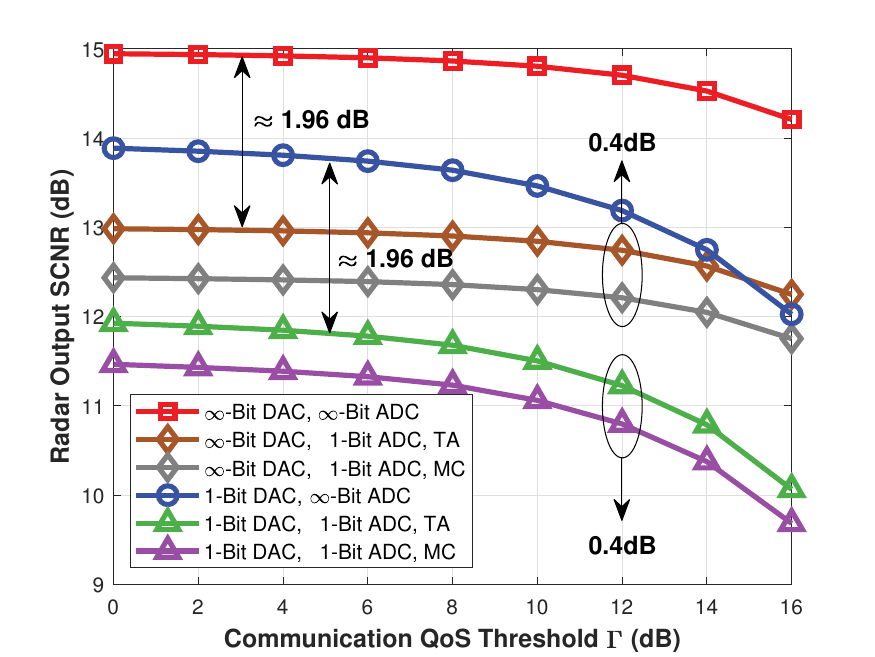}
	\caption{Radar output SCNR versus communication QoS threshold $\Gamma$.}
	\label{fig:exp1}
	\vspace{-1em}
\end{figure}

\textit{Example 1. Impact of Communication QoS Threshold:}
In Fig. \ref{fig:exp1}, we examine the radar output SCNR performance with different communication thresholds $\Gamma$. 
From Fig. \ref{fig:exp1}, we can obtain the following observations:
\textit{1):} A trade-off exists between radar output SCNR and communication QoS requirements. Specifically, as the communication QoS requirement $\Gamma$ increases, the radar output SCNR decreases for all methods. 
This occurs because the radar and communication systems compete for design resources.
\textit{2):} With a small communication QoS threshold $\Gamma$, quantizing with 1-bit DACs results in a 1dB SCNR loss compared to the ideal $\infty$-Bit DACs. 
The performance gap between 1-bit and $\infty$-Bit DACs becomes more pronounced as the communication threshold $\Gamma$ increases.
\textit{3):} Across all communication QoS thresholds $\Gamma$, the loss between DFRC systems with $\infty$-Bit ADCs and 1-bit ADCs is about 1.96dB, which confirms \textit{Insight \ref{ins:1}}. Furthermore, the gap between the theoretical SCNR and the one obtained by MC simulations is only about 0.4dB.

\begin{figure}[!t]
	\centering
	\subfigure[]{
		\includegraphics[width=0.47\linewidth]{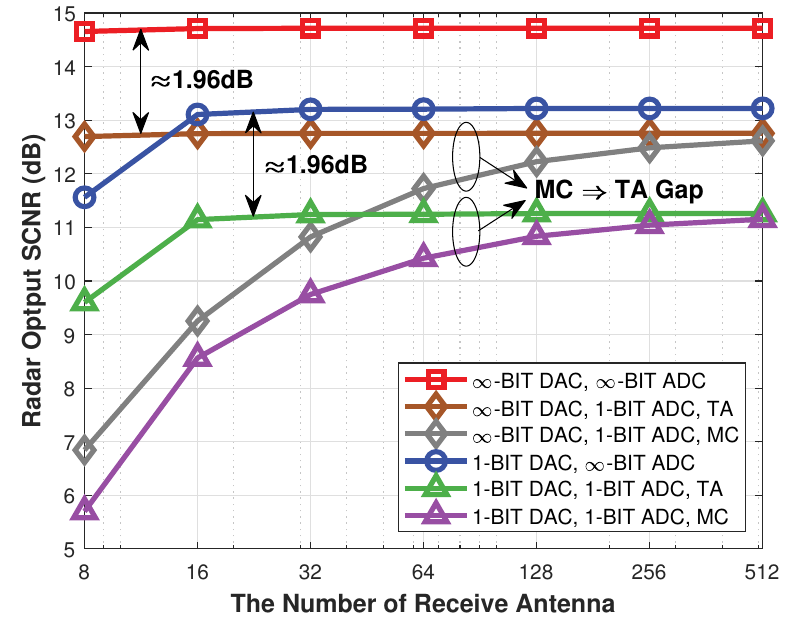}
		\label{fig:exp2a}
	}
	\hspace{-1.em}
	\subfigure[]{
		\includegraphics[width=0.47\linewidth]{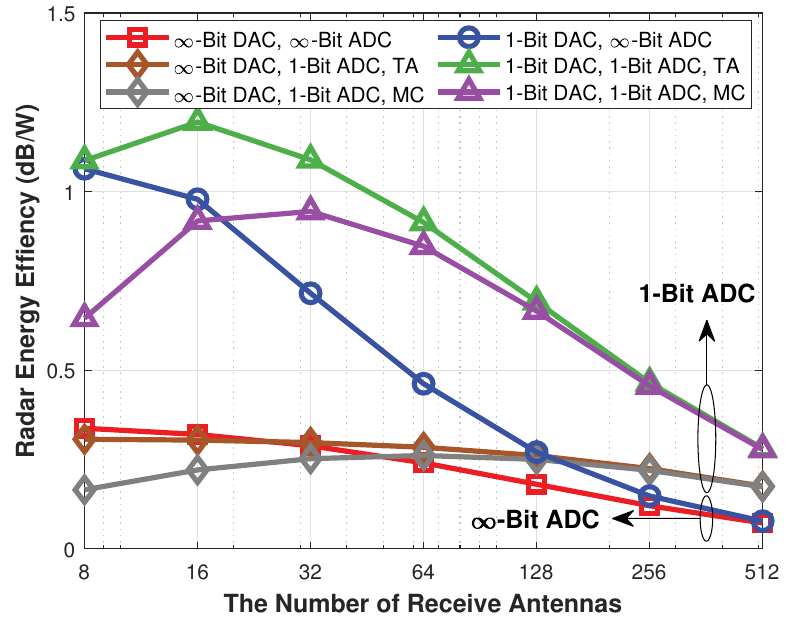}
		\label{fig:exp2b}
	}
	\caption{Radar performance with different $N_{\mathrm{R}}$. (a) Radar output SCNR versus $N_{\mathrm{R}}$. (b) Radar EE versus $N_{\mathrm{R}}$.}
	\label{fig:exp2}
	\vspace{-1em}
\end{figure}
\textit{Example 2. Impact of $N_\mathrm{R}$:}
In Fig. \ref{fig:exp2a}, we present the radar output SCNR versus  $N_{\mathrm{R}}$.
From Fig. \ref{fig:exp2a}, we can obtain the following observations:
\textit{1):} The gap between the TA of 1-bit ADCs and $\infty$-Bit ADCs remains consistently around 1.96dB.
\textit{2):} The gap between the TA and MC for the 1-bit ADCs becomes increasingly negligible as  $N_{\mathrm{R}}$ increases. 
To be more specific, when the number of receive antennas reaches $N_{\mathrm{R}} = 128$, the gap between TA and MC is less than 0.4dB.
Combining the above two observations, we can conclude that the performance analysis and \textit{Insight \ref{ins:2}} presented in Sec. \ref{Sec:3} are accurate.

In Fig. \ref{fig:exp2b}, we present the radar EE versus  $N_{\mathrm{R}}$. 
It is evident that for DFRC with all configurations, increasing the number of receive antennas leads to a degradation in radar EE. 
Furthermore, the DFRC system with $\infty$-Bit ADCs and $\infty$-Bit DACs exhibits the worst radar EE performance, while the DFRC system with 1-bit ADCs and 1-bit DACs outperforms all other DFRC configurations.
These results demonstrate that the DFRC system with 1-bit ADCs and 1-bit DACs is the most energy efficient strategy among the DFRC systems evaluated.

\begin{figure}[!t]
	\centering
	\subfigure[]{
		\includegraphics[width=0.47\linewidth]{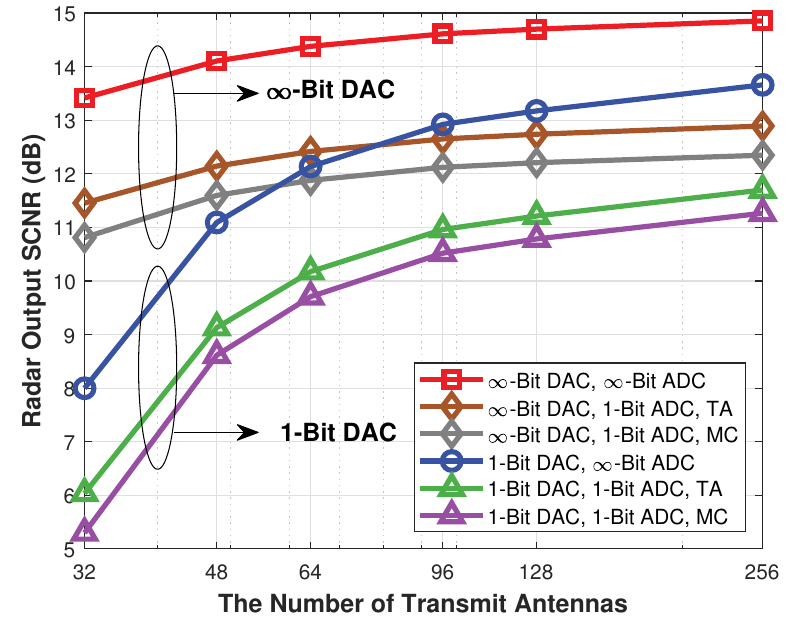}
		\label{fig:exp3a}
	}
	\hspace{-1.em}
	\subfigure[]{
		\includegraphics[width=0.47\linewidth]{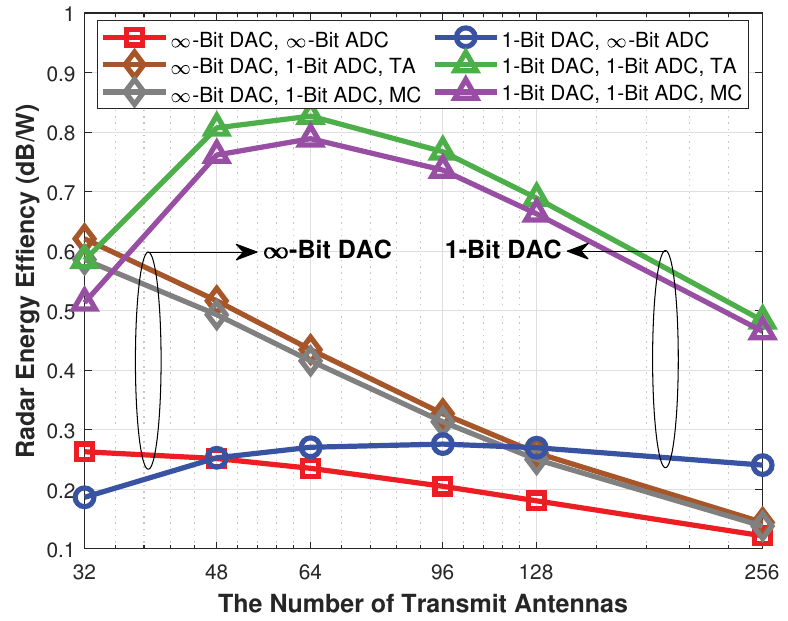}
		\label{fig:exp3b}
	}
	\caption{Radar performance with different $N_{\mathrm{T}}$. (a) Radar output SCNR versus $N_{\mathrm{T}}$. (b) Radar EE versus $N_{\mathrm{T}}$.}
	\label{fig:exp3}
	\vspace{-1em}
\end{figure}
\textit{Example 3. Impact of $N_\mathrm{T}$:}
In Fig. \ref{fig:exp3a}, we study the radar output SCNR versus $N_\mathrm{T}$.
From Fig. \ref{fig:exp3a}, we can obtain the following observations:
\textit{1):} For all DFRC systems, with the increase in $N_\mathrm{T}$, the radar output SCNR firstly improves significantly, and tends to saturate.
This result indicates that a moderately large number of $N_{\mathrm{T}}$ is sufficient to achieve satisfactory performance, while further increasing the number of $N_{\mathrm{T}}$ does not result in further performance improvements, and instead causing significant power consumption and hardware cost.
\textit{2):} Compared with the $\infty$-Bit DACs, the performance for DFRC systems with 1-bit DACs improves more remarkably with the number of the $N_{\mathrm{T}}$.
This is because the quantization by 1-bit DACs results in a discrete sequence $\mathcal{X}_{\text{DAC}}^{\text{1-Bit}}$, which has fewer degrees of freedom and is more sensitive to changes in the number of $N_{\mathrm{T}}$.

In Fig.~\ref{fig:exp3b}, we present the radar EE versus $N_{\mathrm{R}}$. 
Similar to Fig.~\ref{fig:exp2b}, as $N_{\mathrm{T}}$ increases, the radar EE of all DFRC systems tends to decrease, since the total power consumption typically grows linearly with $N_{\mathrm{T}}$, while the radar SCNR gain from adding more antennas becomes less pronounced under the considered constraints and quantization effects. 
Notably, the DFRC system with 1-bit ADC and 1-bit DAC outperforms those with other configurations, validating the effectiveness of adopting this architecture in massive MIMO scenarios. 
Additionally, the DFRC system with 1-bit ADC and 1-bit DAC reaches its peak performance at $N_{\mathrm{T}} = 64$\footnote{We note that $N_{\mathrm{T}}=64$ is optimal only for the specific parameter setting considered here, and the optimal $N_{\mathrm{T}}$ that leads to maximum REE may shift under different scenarios and requirements.}, demonstrating that a moderate number of transmit antennas can effectively balance performance and EE.

\begin{figure}[t]
	\centering
	\includegraphics[width=0.7\linewidth]{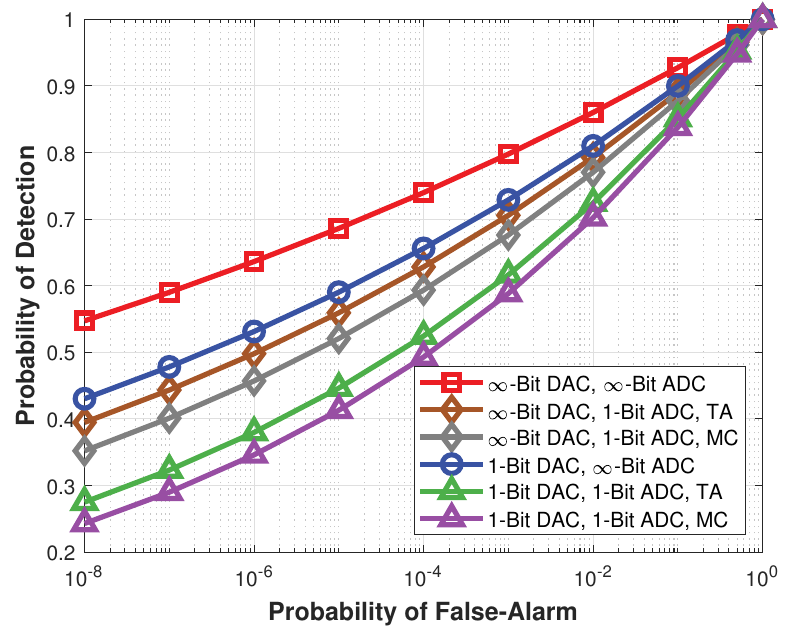}
	\caption{ROC of the DFRC detector with $\Gamma = 12$dB.}
	\label{fig:exp4}
	\vspace{-1em}
\end{figure}
\textit{Example 4. Radar Detection Performance:}
In Fig. \ref{fig:exp4}, we show the radar receiver operating characteristic (ROC) of the DFRC detector proposed in Sec. \ref{Sec:3} with communication QoS threshold $\Gamma = 12$dB.
We observe that the $\mathrm{Pr}_\mathrm{D}$ for all DFRC systems increases with the $\mathrm{Pr}_\mathrm{FA}$.
In addition, the $\mathrm{Pr}_\mathrm{D}$ gap decreases with increasing $\mathrm{Pr}_\mathrm{FA}$.
Furthermore, when jointly considering Figs.\ref{fig:exp1} and \ref{fig:exp4}, it is observed that higher SCNR yields superior detection performance, thereby validating the analysis in Sec.\ref{Sec:3-1}.

\vspace{-0.5em} \subsection{QoD-Constrained 1BitDFRC Design}

In this subsection, we examine the performance of the 1BitDFRC design proposed in Sec. \ref{Sec:5}, where the radar SNCR requirement is constrained to facilitate comparison and validate communication performance.
In the validation simulations, we use 8-PSK modulation with a radar SCNR threshold of $\chi = 12$ dB and $N_{\mathrm{T}} = 32$ transmit antennas.

\begin{figure}[!t]
	\centering
	\subfigure[]{
		\includegraphics[width=0.47\linewidth]{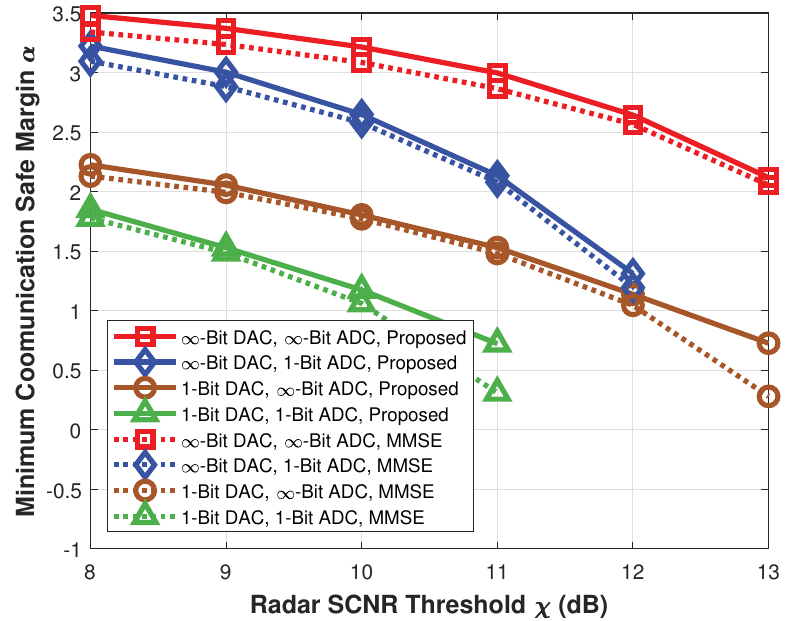}
		\label{fig:exp5a}
	}
	\hspace{-1.em}
	\subfigure[]{
		\includegraphics[width=0.47\linewidth]{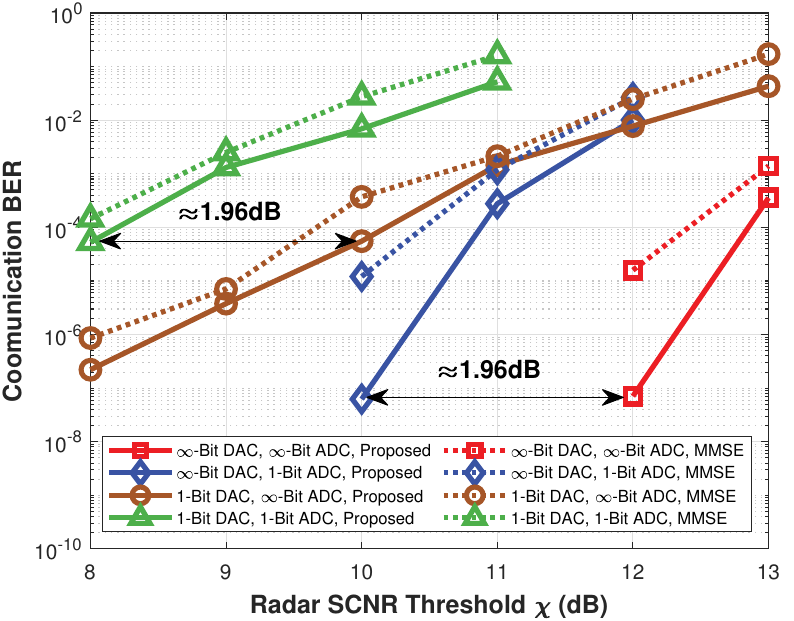}
		\label{fig:exp5b}
	}
	\caption{Communication performance evaluation with $\text{SNR}_\mathrm{C}=15$dB. (a) Minimum communication safe margin $\alpha$ (objective value in \eqref{eq:44}) versus radar SCNR threshold $\chi$. (b) Communication BER versus radar SCNR threshold.}
	\label{fig:exp5}
	\vspace{-1em}
\end{figure}
\textit{Example 5. Impact of Radar SCNR Threshold:}
In Fig. \ref{fig:exp5a}, we study the impact of radar SCNR threshold $\chi$ to communication performance.
From Fig. \ref{fig:exp5a}, we observe that as the radar SCNR threshold increases, the minimum communication safe margin $\alpha$ decreases for all DFRC systems.  
Additionally, across all considered radar SCNR thresholds, the minimum communication safe margin $\alpha$ for the adopted proposed metric is consistently larger than that of the conventional MMSE-based design. 
This result confirms the effectiveness of the proposed safe-margin metric in driving the received signal away from the decision boundary.
As expected, from Fig. \ref{fig:exp5b}, we observe that as the radar SCNR threshold increases, the communication BER also increases for all DFRC systems. 
Additionally, the DFRC systems employing safe margin technology are able to achieve lower BER performance compared to those using the MMSE-based design. 
Fig. \ref{fig:exp5} highlights the advantages and effectiveness of the proposed safe margin-based design in guaranteeing communication BER performance in DFRC.

\begin{figure}[!t]
	\centering
	\subfigure[8-PSK]{
		\includegraphics[width=0.47\linewidth]{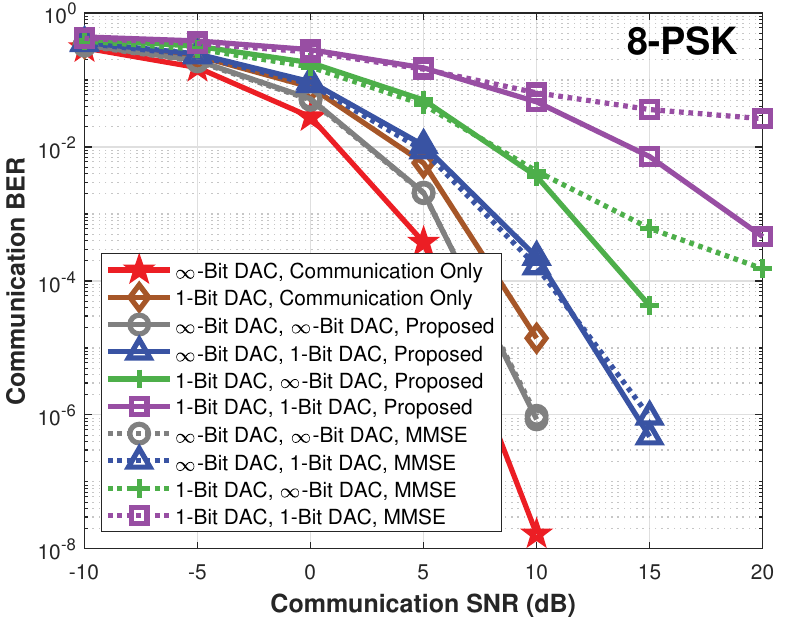}
	}
	\hspace{-1em}
	\subfigure[16-PSK]{
		\includegraphics[width=0.47\linewidth]{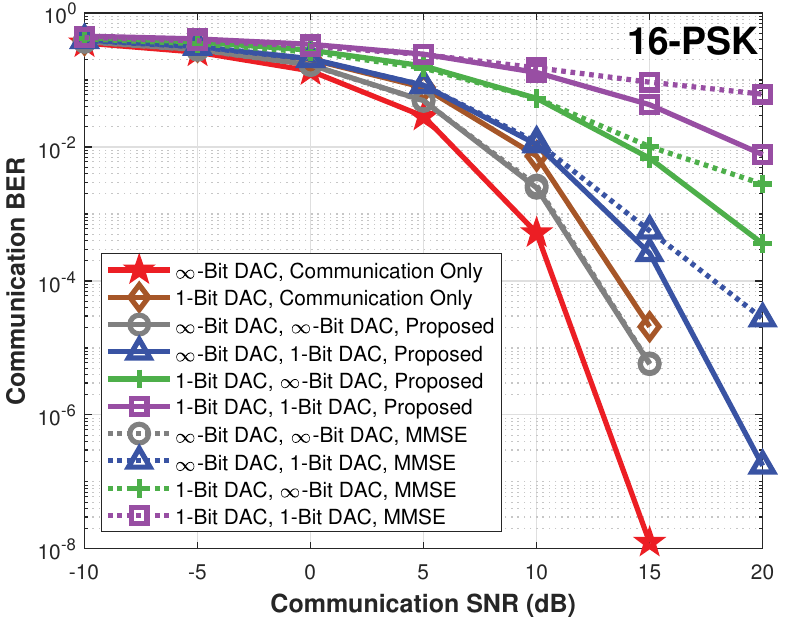}
	}
	\caption{Communication BER versus the communication $\text{SNR}_\mathrm{C}$ with different modulation order $\mathcal{M}$.}
	\label{fig:exp7}
	\vspace{-1em}
\end{figure}
\textit{Example 6. Impact of Modulation Order:}
In Fig. \ref{fig:exp7}, we present the communication BER performance versus the communication $\text{SNR}_\mathrm{C}$ with different PSK modulation order.
From Fig. \ref{fig:exp7}, we obtain the following three observations:
\textit{1):} For all PSK modulation orders and methods, as the communication $\text{SNR}_\mathrm{C}$ gradually increases, the communication BER performance improves.
\textit{2):} When the communication $\text{SNR}_\mathrm{C}$ is 10dB or less, safe margin and MMSE-based DFRC systems can achieve nearly the same communication performance. 
However, when the communication $\text{SNR}_\mathrm{C}$ exceeds 10dB, the safe margin-based design in DFRC systems with a 1-bit architecture can achieve better performance than the MMSE-based design.
\textit{3):} Increasing the PSK modulation order $\mathcal{M}$ leads to worse communication BER performance. This is because higher-order modulation results in smaller PSK decision regions, making the system more sensitive to noise and interference.

\begin{figure}
	\centering
	\includegraphics[width=0.7\linewidth]{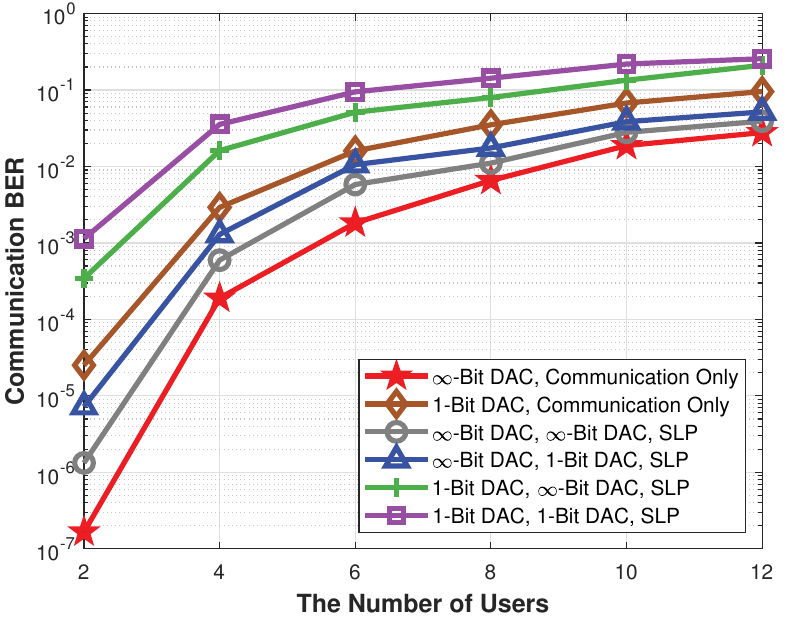}
	\caption{Communication BER versus the number of users $U$.}
	\label{fig:exp_new_7}
	\vspace{-1em}
\end{figure}

\textit{Example 7. Impact of the Number of Users $U$:}
In Fig.~\ref{fig:exp_new_7}, we evaluate the communication BER performance versus the number of downlink users~$U$, under communication SNR of $\text{SNR}_\mathrm{C} = 15$~dB and radar SCNR threshold of $\chi = 8$~dB.
From Fig. \ref{fig:exp_new_7}, it is observed that the communication BER increases as the number of users grows.
This is because supporting more users requires higher spatial degrees of freedom, which are limited by the fixed number of transmit antennas.
Therefore, to accommodate more users while maintaining reliable performance, it may be necessary to increase the number of transmit antennas.

\begin{figure}[t]
	\centering
	\includegraphics[width=0.7\linewidth]{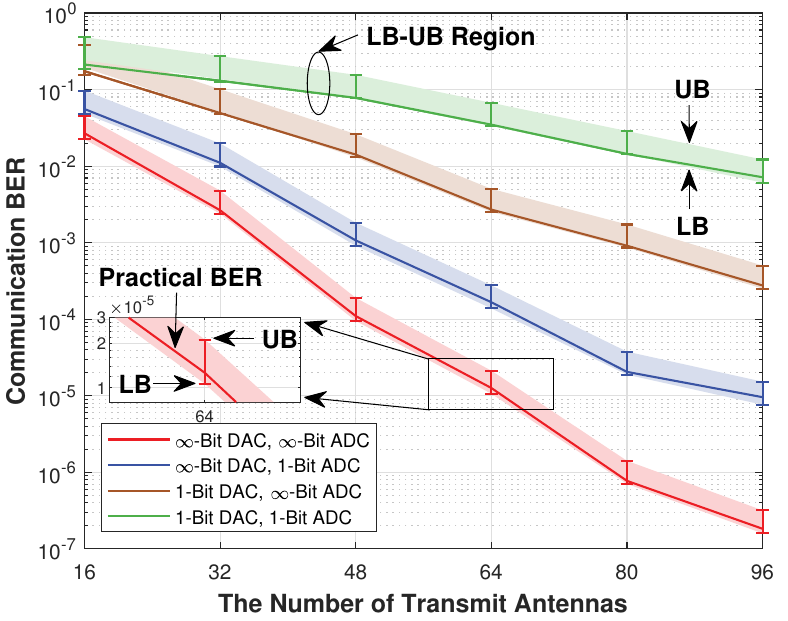}
	\caption{Communication BER versus $N_\mathrm{T}$.}
	\label{fig:exp8}
	\vspace{-1em}
\end{figure}
\textit{Example 8. Validation of SEP boundary:}
In Fig. \ref{fig:exp8}, we show the communication BER versus $N_\mathrm{T}$ with $\text{SNR}_\text{C} = 5$dB.
From Fig. \ref{fig:exp8}, we observe the same trends as in Fig. \ref{fig:exp3a}, where increasing the $N_\mathrm{T}$ leads to more favorable communication BER performance. 
Additionally, for all DFRC systems, the practical BER consistently falls within the lower-bound (LB) and upper-bound (UB) region. 
These results validate \textit{Theorem \ref{the:2}} and verify that maximizing the safe margin is an effective strategy to guarantee communication BER performance.

\vspace{-0.5em} \subsection{Convergence Performance}
In this subsection, we examine the performance of convergence performance of the proposed algorithms.
Parameter settings are the same as those used in the above two subsections.

\begin{table}[t]
	\centering
	\caption{CPU Time (Seconds) Versus $N_\mathrm{R}$}
	\begin{tabular}{cc||llllll}
		\hline
		\multicolumn{2}{c||}{DFRC Configurations}                          & \multicolumn{6}{c}{The Number of Receive Antennas}                                                                                                   \\ \hline
		\multicolumn{1}{c|}{DAC}                           & ADC          & \multicolumn{1}{c|}{16}   & \multicolumn{1}{c|}{32}   & \multicolumn{1}{c|}{64}   & \multicolumn{1}{c|}{128}   & \multicolumn{1}{c|}{256}  & \multicolumn{1}{c}{512} \\ \hline
		\multicolumn{1}{c|}{\multirow{2}{*}{$\infty$-Bit}} & $\infty$-Bit & \multicolumn{1}{l|}{1.08} & \multicolumn{1}{l|}{1.43} & \multicolumn{1}{l|}{1.55} & \multicolumn{1}{l|}{1.88} & \multicolumn{1}{l|}{1.97} & 2.26                    \\ \cline{2-8} 
		\multicolumn{1}{c|}{}                              & 1-Bit        & \multicolumn{1}{l|}{1.38} & \multicolumn{1}{l|}{1.61} & \multicolumn{1}{l|}{1.67} & \multicolumn{1}{l|}{1.89} & \multicolumn{1}{l|}{1.98} & 2.27                    \\ \hline
		\multicolumn{1}{c|}{\multirow{2}{*}{1-Bit}}        & $\infty$-Bit & \multicolumn{1}{l|}{5.17} & \multicolumn{1}{l|}{5.19} & \multicolumn{1}{l|}{4.86} & \multicolumn{1}{l|}{6.21} & \multicolumn{1}{l|}{6.73} & 6.89                    \\ \cline{2-8} 
		\multicolumn{1}{c|}{}                              & 1-Bit        & \multicolumn{1}{l|}{4.91} & \multicolumn{1}{l|}{4.96} & \multicolumn{1}{l|}{5.00} & \multicolumn{1}{l|}{6.55} & \multicolumn{1}{l|}{6.75} & 7.10                    \\ \hline
	\end{tabular}
	\label{tab:1}
	\vspace{-1em}
\end{table}

\begin{table}[t]
	\centering
	\caption{CPU Time (Seconds) Versus  $N_\mathrm{T}$}
	\begin{tabular}{cc||llllll}
		\hline
		\multicolumn{2}{c||}{DFRC Configurations}                          & \multicolumn{6}{c}{The Number of Transmit Antennas}                                                                                                   \\ \hline
		\multicolumn{1}{c|}{DAC}                           & ADC          & \multicolumn{1}{c|}{32}   & \multicolumn{1}{c|}{48}   & \multicolumn{1}{c|}{64}   & \multicolumn{1}{c|}{96}   & \multicolumn{1}{c|}{128}  & \multicolumn{1}{c}{256} \\ \hline
		\multicolumn{1}{c|}{\multirow{2}{*}{$\infty$-Bit}} & $\infty$-Bit & \multicolumn{1}{l|}{0.64} & \multicolumn{1}{l|}{0.64} & \multicolumn{1}{l|}{0.64} & \multicolumn{1}{l|}{0.64} & \multicolumn{1}{l|}{0.65} & 0.67                    \\ \cline{2-8} 
		\multicolumn{1}{c|}{}                              & 1-Bit        & \multicolumn{1}{l|}{0.65} & \multicolumn{1}{l|}{0.64} & \multicolumn{1}{l|}{0.64} & \multicolumn{1}{l|}{0.65} & \multicolumn{1}{l|}{0.65} & 0.67                    \\ \hline
		\multicolumn{1}{c|}{\multirow{2}{*}{1-Bit}}        & $\infty$-Bit & \multicolumn{1}{l|}{0.56} & \multicolumn{1}{l|}{1.04} & \multicolumn{1}{l|}{2.86} & \multicolumn{1}{l|}{4.61} & \multicolumn{1}{l|}{5.48} & 6.92                    \\ \cline{2-8} 
		\multicolumn{1}{c|}{}                              & 1-Bit        & \multicolumn{1}{l|}{0.55} & \multicolumn{1}{l|}{1.02} & \multicolumn{1}{l|}{2.55} & \multicolumn{1}{l|}{4.75} & \multicolumn{1}{l|}{5.64} & 7.02                    \\ \hline
	\end{tabular}
	\label{tab:2}
	\vspace{-1em}
\end{table}

\textit{Example 9. Comparison of Computational Efficiency:}
Tables \ref{tab:1} and \ref{tab:2} present the CPU time (in seconds) versus $N_{\mathrm{R}}$ and transmit antennas $N_{\mathrm{T}}$ for QoS-constrained design, respectively.
As expected, we observe that as the number of receive antennas $N_{\mathrm{R}}$ and transmit antennas $N_{\mathrm{T}}$ increase, the CPU time required also increases. 
Compared with DFRC systems with $\infty$-Bit DACs, DFRC systems with 1-bit DACs take more time to converge. 
This increased time is due to the binary optimization required by the 1-bit ADC, which is computationally intensive.
Additionally, we find that DFRC systems with either 1-bit or $\infty$-Bit DACs does not impact the convergence time significantly.
Furthermore, the proposed method can converge within 10 seconds, which verifies the computational efficiency of the proposed approach.

\begin{figure}[t]
	\centering
	\includegraphics[width=0.7\linewidth]{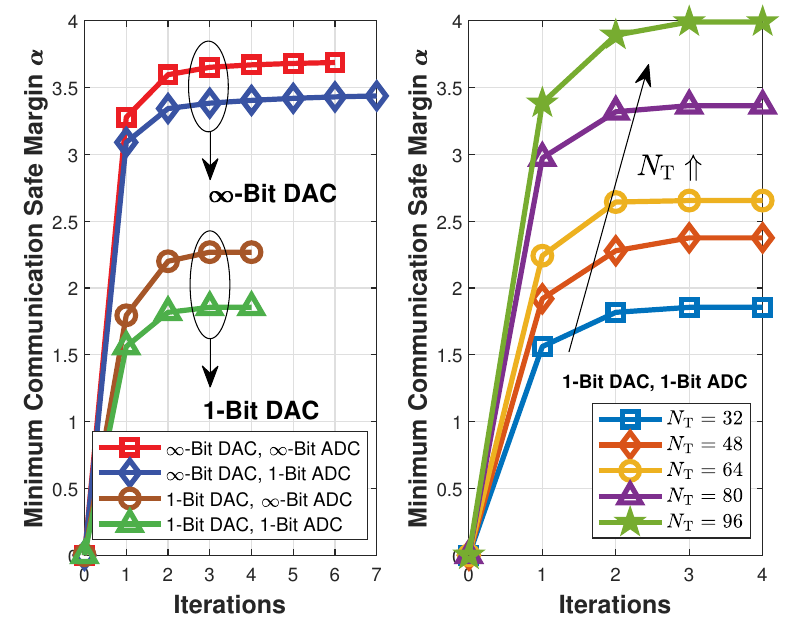}
	\caption{Convergence property of all DFRC systems with $\chi=8$dB.}
	\label{fig:exp10}
	\vspace{-1em}
\end{figure}
\textit{Example 10. Comparison of Convergence Efficiency:}
In Fig. \ref{fig:exp10}, we present the convergence behavior of the proposed solutions for the QoD-constrained design.
From the left side of Fig. \ref{fig:exp10}, it is evident that for all DFRC systems, the objective function becomes stationary after several iterations, which verifies the convergence efficiency of the proposed algorithm. 
Additionally, we observe similar trends where $\infty$-Bit DACs provide more degrees of freedom, allowing convergence to larger values, while 1-bit DACs converge to lower values.
From the right side of Fig. \ref{fig:exp10}, as expected, increasing $N_{\mathrm{T}}$ leads to convergence to higher values, which aligns with the results observed in Fig. \ref{fig:exp8}. 
Furthermore, for DFRC systems with different numbers of transmit antennas $N_{\mathrm{T}}$, the proposed algorithm can always converge within 10 iterations, demonstrating again the efficiency of the proposed algorithm.

\vspace{-0.5em} \subsection{Comparison with the State-of-the-Art}
\begin{table}[t]
	\centering
	\caption{Comparison between proposed algorithm and ADMM method.}
	\begin{tabular}{ccclcc}
		\hline
		\multirow{2}{*}{$N_{\mathrm{T}}$} & \multicolumn{2}{c}{Radar Output SCNR (dB)} &  & \multicolumn{2}{c}{CPU Time (Seconds)} \\ \cline{2-3} \cline{5-6} 
		&  Proposed & AO \cite{wang2023joint} &  & Proposed & AO \cite{wang2023joint} \\ \hline
		32  & 5.30   & 5.18  &  & 0.55  & 8.68  \\
		48  & 8.62   & 8.64  &  & 1.02  & 10.51 \\
		64  & 9.70   & 9.51  &  & 2.55  & 11.84 \\
		96  & 10.51  & 10.31 &  & 4.75  & 14.71  \\
		128 & 10.78  & 10.70 &  & 5.64  & 20.60 \\
%		256 & 11.26  & 11.26 &  & 7.02  &  \\
 		\hline
	\end{tabular}
	\label{tab_3}
\end{table}

In this subsection, we compare the proposed algorithm with a baseline benchmark. 
Specifically, the alternating direction method of multipliers (ADMM) based algorithm developed in \cite{wang2023joint}, and subsequently adopted in the recent work \cite{STCOM2026}, is selected as the primary benchmark.

\textit{Example 11. Comparison with ADMM-based Algorithm:}
In Tab. \ref{tab_3}, we compare our proposed algorithm with the ADMM based method. 
Since both methods employ the same receiver side design, we focus on comparing the impact of the transmit antenna configuration.
Regarding the radar output SCNR, the proposed algorithm outperforms the ADMM based method. 
This is because the ADMM based approach converges to a suboptimal local minimum in non-convex problems. 
In contrast, our proposed BnB based method efficiently searches the solution space to achieve a near-global optimal solution.

The proposed algorithm also maintains a lower CPU time than the ADMM based method. 
This efficiency stems from the fact that the ADMM based method in \cite{wang2023joint} employs a double-layer optimization structure, requiring numerous inner iterations to update the transmit signal $\bm{x}$ under 1-bit DAC constraints, followed by multiple outer iterations. 
Each inner iteration involves computationally expensive operations, such as matrix inversion and SVD of an $N_\mathrm{T} \times N_\mathrm{T}$ matrix. 
In contrast, our proposed method utilizes a single-layer iteration and avoids these high-complexity operations. 
As $N_{\mathrm{T}}$ increases, the computational burden of the ADMM based method becomes more pronounced, leading to a significant escalation in CPU time.

\begin{figure}
	\centering
	\includegraphics[width=0.7\linewidth]{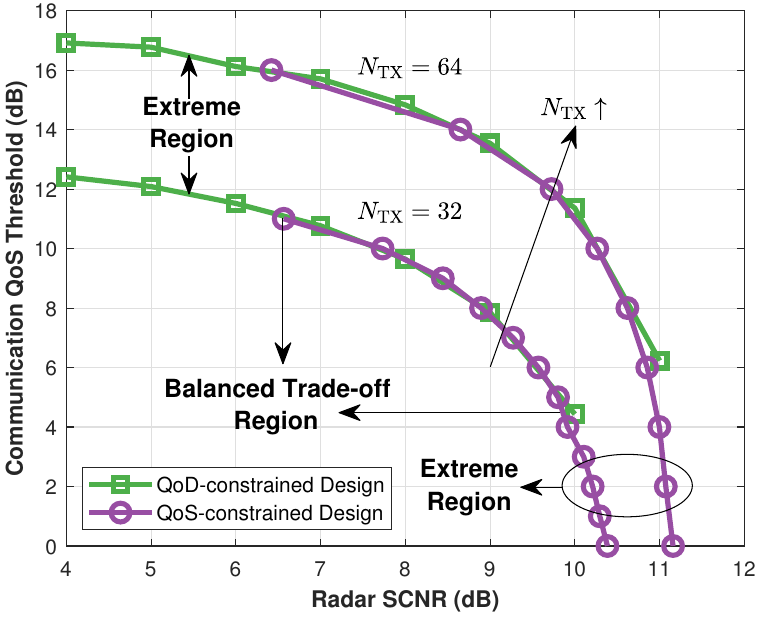}
	\caption{{Pareto frontiers of the QoS- and QoD-constrained 1BitDFRC designs.}}
	\label{fig:new_pareto}
	\vspace{-1em}
\end{figure}

\vspace{-0.5em} 
\subsection{Comparison Between QoS- and QoD-Constrained Designs}
As discussed in Remark \ref{New_Rem_2}, the QoS-constrained and QoD-constrained formulations constitute two complementary approaches to 1BitDFRC design, attaining essentially identical performance while offering distinct advantages in different operating regimes. To support this claim, we characterize the Pareto frontiers of both designs in the following example.

\textit{Example 12 (Pareto Frontier of 1BitDFRC):} Fig.~\ref{fig:new_pareto} depicts the Pareto frontiers achieved by the QoS- and QoD-constrained designs.
A pronounced trade-off between radar and communication performance is observed, which is consistent with the behavior reported in Figs.~\ref{fig:exp1} and \ref{fig:exp5}.
More importantly, the two Pareto frontiers are nearly indistinguishable in the balanced trade-off region, which supports Remark~\ref{New_Rem_2}.
Differences appear only in the two extreme regions.
Specifically, the QoS-constrained design achieves better radar performance when the communication QoS requirement is mild, whereas the QoD-constrained design yields better communication performance when the radar SCNR requirement is mild.
This behavior can be explained by the feasibility properties of two formulations.
The QoS-constrained problem becomes difficult to keep feasible under stringent communication QoS requirements, and the same holds for the QoD-constrained problem under stringent radar SCNR requirements.
The two designs are therefore complementary, and the choice between them should depend on the specific requirements and constraints of the application scenario.
Moreover, as $N_{\mathrm{TX}}$ increases, the Pareto frontier shifts outward toward the upper-right corner, indicating simultaneous improvement in both radar and communication performance.
This is also consistent with the observation in Fig.~\ref{fig:exp3}.

\vspace{-0.5em} 
\subsection{Extension to Fairness-aware Multi-Target 1BitDFRC}
In this subsection, we examine the performance of 1-bit DFRC in multi-target scenarios. 
Since the core performance metrics have been revealed and studied in the previous section, we now focus on more visual performance metric, i.e., beampattern.

\begin{figure}[t]
	\centering
	\includegraphics[width=0.7\linewidth]{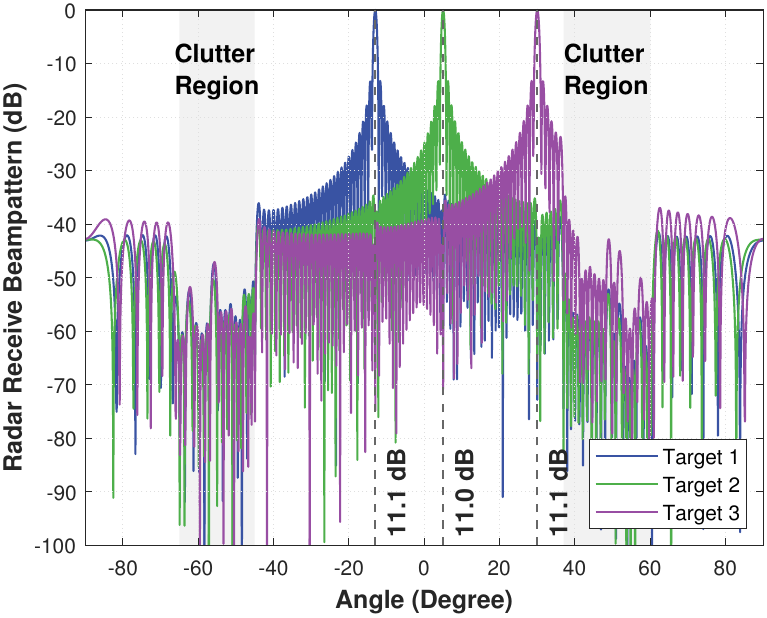}
	\caption{1BitDFRC Receive Beampattern.}
	\label{fig:exp11}
	\vspace{-1em}
\end{figure}

\textit{Example 13. 1BitDFRC Receive Beampattern:}
Fig. \ref{fig:exp11} shows the receive beampattern of the proposed 1BitISAC, where the beampattern of the $k$-th target is computed as $\mathcal{P}_{\mathrm{RX}} (\theta) = | \mathbm{f}_k \mathcal{Q}_{\text{ADC}}^{\text{1-Bit}}  (\mathbm{g}_{\mathrm{R}} \mathbm{g}_{\mathrm{T}}^T \mathbm{x}) |^2 $.
The three targets are located at angles $\{\theta_{\mathrm{T}, 1}, \theta_{\mathrm{T}, 2}, \theta_{\mathrm{T}, 3}\} = \{-13^\circ, 5^\circ, 30^\circ\}$, while the clutter regions are defined as $[-65^\circ, -45^\circ] \cup [37^\circ, 60^\circ]$.
For each receive beampattern, it is clear that it achieves a high peak at the corresponding target while maintaining nulls in the clutter regions (gray regions), ensuring a high detection probability. 
Besides, when detecting Target 1, the receive beampattern maintains a low level at the locations of other targets to ensure no cross-interference, where similar observations can be made for the other targets.
As marked at the bottom of Fig. \ref{fig:exp11}, benefiting from the fairness-aware design, all targets achieve nearly the same output SCNR, thereby ensuring fair detection performance.

\section{Conclusion}\label{Sec:7}

Joint transceiver design for an 1BitDFRC system with massive MIMO is investigated. Two 1BitDFRC designs, i.e., QoS-constrained 1BitDFRC design and QoD-constrained 1BitDFRC design, were considered.
To facilitate the design of the 1BitDFRC system, we first analyze the radar detection performance under 1-bit ADC quantization.
Subsequently, the communication BER performance was also analyzed, providing a new design perspective for measuring the communication performance in the 1BitDFRC sysrem. 
Based on insights from these analyses, the design problems were reformulated and solved. Numerous simulation results were provided, revealing the following insights:
\begin{itemize}
	\item After 1-bit ADC quantization, the proposed 1BitDFRC system suffers from roughly a 1.96 dB performance loss, aligning with the performance analysis.
	These results highlight the feasibility of adopting 1-bit ADCs in massive MIMO DFRC systems, with an predictable and acceptable level of performance loss.
	\item Compared with the conventional communication MMSE, the CI-based SLP method achieves better communication BER performance.
	These results verifies the BER analysis and supports the new design perspective in 1BitDFRC.
	\item Compared with other DFRC configurations, the proposed 1BitDFRC achieves the best EE performance in massive MIMO DFRC systems, albeit at the cost of slight performance sacrifices. 
	These results indicate that the proposed 1BitDFRC successfully balances the trade-off between EE and DFRC performance.
\end{itemize}

Future work may consider extending this study in several directions, including:
\textit{(i)} practical hardware implementation of the proposed 1BitDFRC architecture under hardware impairments (e.g., synchronization errors and analog front-end nonlinearities),
\textit{(ii)} distributed 1BitDFRC systems with collaborative sensing and communication,
\textit{(iii)} more robust and tractable SEP analysis under practical channel and noise conditions,
\textit{(iv)} estimation-aware 1BitDFRC frameworks that jointly account for parameter estimation performance,
and \textit{(v)} multi-bit DFRC performance analysis and design.

\appendices

\section{Proof of Theorem \ref{the:1}}\label{proof:1}

Based on our prior work in \cite[Proposition 1]{deng2022receive}, the mean and variance of $\widetilde{\bm{r}}$ is given by
\begin{equation}\label{neq:45}
	\mathbb{E}\{ \widetilde{\mathbm{r}} | \mathcal{H}_i , \mathbm{\gamma}_i \} =  \frac{2}{\sqrt{\pi\sigma_\mathrm{R}^2} } \mathbm{r}_i, \quad 
	\mathbb{D}\{ \widetilde{\mathbm{r}} | \mathcal{H}_i , \mathbm{\gamma}_i \} = 2 \mathbm{I}_{N_{\mathrm{R}}} .
\end{equation}

In our previous work \cite{deng2022receive}, we assumed that the terms $[\mathbm{f}]_n^* [\widetilde{\mathbm{r}}]_n$ are i.i.d.
Under this assumption, the classical CLT \cite{billingsley2013convergence,williams1991probability,billingsley2008probability} can be applied to approximate the distribution of $z$.
However, in practice, the components of $\widetilde{\mathbm{r}}$ are indeed independent, but not identically distributed (see \eqref{neq:45}).
As a result, the i.i.d. assumption no longer holds, and the classical CLT becomes inapplicable \cite{billingsley2013convergence,williams1991probability,billingsley2008probability}.
To fix this point, we conduct a more rigorous and general analysis by employing the Lyapunov CLT \cite{billingsley2013convergence,williams1991probability,billingsley2008probability}, which accommodates sums of independent but non-identically distributed random variables.
Specifically, we decompose $z$ into a deterministic term and a random term as follows
\begin{equation}
	\begin{aligned}
		z = & {\mathbm{f}}^H\widetilde{\mathbm{r}} = \sum_{n=1}^{N_{\mathrm{R}}} [\mathbm{f}]_n^* [\widetilde{\mathbm{r}}]_n \\
		= & \sum_{n=1}^{N_{\mathrm{R}}} \underbrace{ [\mathbm{f}]_n^* \left( [\widetilde{\mathbm{r}}]_n - \frac{2[\mathbm{r}]_n}{ \sqrt{\pi \sigma_{\mathrm{R}}^2} } \right) }_{\mathcal{A}_n}
		+ \sum_{n=1}^{N_{\mathrm{R}}} \underbrace{ [\mathbm{f}]_n^* \frac{2[\mathbm{r}]_n}{ \sqrt{\pi \sigma_{\mathrm{R}}^2} }}_{\mathcal{D}_n} .
	\end{aligned}
\end{equation}
Note that $\mathcal{A}_n$ is a zero-mean complex random variable, and the variance of $\mathcal{A}_n$ is
\begin{equation}
	\mathbb{D} \{ \mathcal{A}_n | \mathcal{H}_i , \mathbm{\gamma}_i \} = \left| [\mathbm{f}]_n \right|^2 \mathbb{D} \{ [\widetilde{\mathbm{r}}]_n | \mathcal{H}_i , \mathbm{\gamma}_i \} = 2 \left| [\mathbm{f}]_n \right|^2 .
\end{equation}

Define the total variance of $\widetilde{z} = \sum_{n=1}^{N_{\mathrm{R}}} \mathcal{A}_n $ as
\begin{equation}\label{neq:48}
	\beta^2 = \sum_{n=1}^{N_{\mathrm{R}}} \mathbb{D} \{ \mathcal{A}_n | \mathcal{H}_i , \mathbm{\gamma}_i \} = 2 \sum_{n=1}^{N_{\mathrm{R}}} \left| [\mathbm{f}]_n \right|^2 .
\end{equation}

To invoke the complex Lyapunov CLT, we need to verify the Lyapunov condition \cite{billingsley2008probability} for some $\varphi > 0$
\begin{equation}\label{neq:49}
	\lim\limits_{N_{\mathrm{R}} \to \infty} \frac{1}{\beta^{2+\varphi}} \sum_{n=1}^{N_{\mathrm{R}}} \mathbb{E} \left[ | \mathcal{A}_n |^{2 + \varphi} \right] = 0 .
\end{equation}

We fix $\varphi = 2$, so that
\begin{equation}\label{neq:50}
	\begin{aligned}
		\mathbb{E} \left[ | \mathcal{A}_n |^{2 + \varphi} \right]
		& = \mathbb{E} \left[ \left| [\mathbm{f}]_n^* \left( [\widetilde{\mathbm{r}}]_n - \frac{2[\mathbm{r}]_n}{ \sqrt{\pi \sigma_{\mathrm{R}}^2} } \right) \right|^{2 + \varphi} \right] \\
		& \mathop = \limits^{\text{(a)}} \left| [\mathbm{f}]_n \right|^{2 + \varphi} \mathbb{E} \left[ \left| [\widetilde{\mathbm{r}}]_n - \frac{2[\mathbm{r}]_n}{ \sqrt{\pi \sigma_{\mathrm{R}}^2} } \right|^{2 + \varphi} \right] \\
		& = 8 \left| [\mathbm{f}]_n \right|^{4}  ,
	\end{aligned}
\end{equation}
where (a) holds since $ [\widetilde{\mathbm{r}}]_n - {2[\mathbm{r}]_n} / { \sqrt{\pi \sigma_{\mathrm{R}}^2} } \sim \mathcal{CN}(0 , 2) $, such that fourth-order moment is $\mathbb{E} [ | [\widetilde{\mathbm{r}}]_n - {2[\mathbm{r}]_n} / { \sqrt{\pi \sigma_{\mathrm{R}}^2} } |^{4} ] = 8$.

Substituting \eqref{neq:48} and \eqref{neq:50} into \eqref{neq:49}, we obtain
\begin{equation}
	\lim\limits_{N_{\mathrm{R}} \to \infty} \frac{1}{\beta^{2+\varphi}} \sum_{n=1}^{N_{\mathrm{R}}} \mathbb{E} \left[ | \mathcal{A}_n |^{2 + \varphi} \right] = \frac{ 8 \sum_{n=1}^{N_{\mathrm{R}}}\left| [\mathbm{f}]_n \right|^{4}  }{ \left( 2 \sum_{n=1}^{N_{\mathrm{R}}} \left| [\mathbm{f}]_n \right|^2 \right)^2 } .
\end{equation}
Applying the standard inequalities
\begin{align}
	\sum_{n=1}^{N_{\mathrm{R}}}\left| [\mathbm{f}]_n \right|^{4} & \le N_{\mathrm{R}} \max_n \left| [\mathbm{f}]_n \right|^{4} , \\
	\left( \sum_{n=1}^{N_{\mathrm{R}}} \left| [\mathbm{f}]_n \right|^2 \right)^2 & \ge N_{\mathrm{R}}^2 \left( \min_n \left| [\mathbm{f}]_n \right|^2 \right)^2 ,
\end{align}
we obtain the following upper bound
\begin{equation}
	\begin{aligned}
		\lim\limits_{N_{\mathrm{R}} \to \infty} \frac{1}{\beta^{2+\varphi}} \sum_{n=1}^{N_{\mathrm{R}}} \mathbb{E} \left[ | \mathcal{A}_n |^{2 + \varphi} \right]
		\le \frac{2 \max_n \left| [\mathbm{f}]_n \right|^{4}  }{N_{\mathrm{R}} \left( \min_n \left| [\mathbm{f}]_n \right|^2 \right)^2} .
	\end{aligned}
\end{equation}

Since $\mathbm{f}$ is radar receiver filter and both the maximum and minimum entries are bounded in magnitude, leading to $\frac{ \max_n \left| [\mathbm{f}]_n \right|^{4}  }{\left( \min_n \left| [\mathbm{f}]_n \right|^2 \right)^2} \le \mathcal{C} < + \infty$.
Hence, we conclude
\begin{equation}
	\begin{aligned}
		\lim\limits_{N_{\mathrm{R}} \to \infty} \frac{1}{\beta^{2+\varphi}} \sum_{n=1}^{N_{\mathrm{R}}} \mathbb{E} \left[ | \mathcal{A}_n |^{2 + \varphi} \right]
		\sim \mathcal{O}(N_{\mathrm{R}}^{-1}) \to 0 .
	\end{aligned}
\end{equation}

Thus, when $N_{\mathrm{R}} \to \infty$, the Lyapunov condition is satisfied. By the complex Lyapunov CLT, we conclude that the distribution of $z$ converges to a complex Gaussian random variable
\begin{equation}
	\mathbb{E}\{ z | \mathcal{H}_i , \mathbm{\gamma}_i \} = \frac{2}{\sqrt{ \pi\sigma_\mathrm{R}^2} } {\mathbm{f}}^H\mathbm{r}_i, \;\;\; 
	\mathbb{D}\{ z | \mathcal{H}_i , \mathbm{\gamma}_i \} = 2 \| {\mathbm{f}} \|_F^2 .
	\label{S3_eq:proof4}
\end{equation}

Therefore, the conditional distribution of $z$ is given by
\begin{equation}
	\text{Pr} \left( z | \mathcal{H}_i , \mathbm{\gamma}_i \right) = \frac{1}{2\pi \left\| {\mathbm{f}} \right\|_F^2 } \exp \Bigg( - \frac{ \big| z - \frac{2}{\sqrt{\pi \sigma_\mathrm{R}^2}} {\mathbm{f}}^H \mathbm{r}_i \big|^2 }{ 2 \left\| {\mathbm{f}} \right\|_F^2 } \Bigg) .
\end{equation}

\section{Proof of Theorem \ref{the:3}}\label{proof:3}
To prove the convergence of algorithm \ref{alg:1}, it is sufficient to illustrate the following two issues.
\begin{enumerate}
	\item Monotonicity: Based on the property of MM, it follows that
	\begin{equation}
		\begin{aligned}
			{ \text{QSCNR} } & ({\mathbm x}_t) = \overline{ \text{QSCNR} } ({\mathbm x}_t | {\mathbm{x}}_t) \\
			& \le \overline{ \text{QSCNR} } ({\mathbm x}_{t+1} | {\mathbm{x}}_t) \le { \text{QSCNR} } ({\mathbm x}_{t+1}) .
		\end{aligned}
	\end{equation}
	As the index $t$ increases, $\overline{ \text{QSCNR} } ({\mathbm x} | {\mathbm{x}}_t)$ is monotonically improving.
	\item Boundedness: The boundedness of $\text{QSCNR} \left( \mathbm{x} \right)$ can be showed by following inequalities
	\begin{equation}
		\begin{aligned}
			& \text{QSCNR} \left( \mathbm{x} \right)  \mathop  \le \limits^{(a)} \text{QSCNR}_{wo} \left(  \mathbm{x} \right) \\
			& = \frac{2}{\pi} \mathrm{SNR}_\mathrm{R} {{{\mathbm{x}}^H}{\mathbm{G}}_0^H {{\mathbm{G}}_0}{\mathbm{x}}} = \frac{2}{\pi} \mathrm{SNR}_\mathrm{R} \| \mathbm{g}_\mathrm{T}^T(\theta_0) {\mathbm x}\|_F^2 \\
			& \mathop  \le \limits^{(b)} \frac{2E}{\pi} \mathrm{SNR}_\mathrm{R}.
		\end{aligned}
	\end{equation}
	where $(a)$ is because the $\text{QSCNR} \left( \mathbm{x} \right)$ in the presence of clutter sources is always lower than in scenarios without clutter sources $\text{QSCNR}_{wo} \left( \mathbm{x} \right)$.
	$(b)$ is because the optimal solution of $\max_{\mathbm{x}} \| \mathbm{g}_\mathrm{T}^T(\theta_0) {\mathbm x}\|_F^2 $ is $\mathbm{x} = \sqrt{E} \mathbm{g}_T^*(\theta_0)$ with optimal value $E$.
	Since the transmit power $E$ is limited, the $\text{QSCNR} \left( \mathbm{x} \right)$ is always bounded.
\end{enumerate}
Based on above two properties, we can conclude the sequence $\{ \text{QSCNR}(\mathbm{x}_t) \}$ is non-decreasing and converges to a local maximum.

\section{Solution to Problem \eqref{eq_42_new}}\label{New_App_4}
For ease of readability, we reproduce \eqref{eq_42_new} below
\begin{subequations}\label{app_eq_42_new}
	\begin{align}
		\mathop {\max }\limits_{{\mathbm{x}},\{\mathbm{f}_k\}} \quad & \min_{k} \text{QSCNR}_k \left(\{\mathbm{f}_k\} , \mathbm{x}\right) \\
		{\text{s}}.{\text{t}}. \quad & \min \{ \Re ( \kappa_1 \bar{\mathbm{h}}_u^H \mathbm{x} ) , \Re ( \kappa_2 \bar{\mathbm{h}}_u^H \mathbm{x} )  \} \ge \lambda_u , \forall u , \\
		& {\mathbm{x}} \in \mathcal{X}_{\text{DAC}}^{\text{1-Bit}} = \left\{ \frac{E}{{\sqrt {2{M_{\text{T}}}} }}  ({ \pm 1 \pm \jmath}) \right\} .
	\end{align}
\end{subequations}

By introducing an auxiliary variable $\rho$, we equivalently reformulate \eqref{app_eq_42_new} as
\begin{subequations}\label{eq_57}
	\begin{align}
		\mathop {\max }\limits_{{\mathbm{x}},\{\mathbm{f}_k\}, \rho} \quad & \rho \label{eq_57_a}\\
		{\text{s}}.{\text{t}}. \quad\; & \min \{ \Re ( \kappa_1 \bar{\mathbm{h}}_u^H \mathbm{x} ) , \Re ( \kappa_2 \bar{\mathbm{h}}_u^H \mathbm{x} )  \} \ge \lambda_u , \forall u , \label{eq_57_b} \\
		& \text{QSCNR}_k \left(\{\mathbm{f}_k\} , \mathbm{x}\right) \ge \rho, \forall k , \label{eq_57_d} \\
		& {\mathbm{x}} \in \mathcal{X}_{\text{DAC}}^{\text{1-Bit}} = \left\{ \frac{E}{{\sqrt {2{M_{\text{T}}}} }}  ({ \pm 1 \pm \jmath}) \right\} . \label{eq_57_c}
	\end{align}
\end{subequations}

Constraint \eqref{eq_57_b} can be handled in a same manner via \eqref{NEQ_29_new} in Sec. \ref{Sec_4_B}, which gives
\begin{equation}
	\begin{aligned}
		\eqref{eq_57_b} & \Leftrightarrow \left\{
		\Re \left\{ \kappa_1 \bar{\mathbm{h}}_u^H \mathbm{x} \right\} \ge \lambda_u  , 
		\Re \left\{ \kappa_2 \bar{\mathbm{h}}_u^H \mathbm{x} \right\} \ge \lambda_u 
		\right\}  ,  \\ 
		& \Leftrightarrow \Re \left\{ \mathbm{A}_u \mathbm{x} \right\} \ge \lambda_u  \mathbm{1}_2 = \mathbm{\lambda}_u .
	\end{aligned}
\end{equation}

Constraint \eqref{eq_57_d} can be handled in a same manner via \eqref{eq:34} and \eqref{eq:37} in Sec. \ref{Sec_4_B}, which gives
\begin{subequations}
	\begin{align}
		& \text{QSCNR}_k({\mathbm x}) \ge
		\Re\left\{ \mathbm{w}_{t,k}^H \mathbm{x} \right\} + \text{const}_{2,k} ,
	\end{align}
\end{subequations}
where
\begin{equation}
	\mathbm{w}_{t,k} = 2 \mu \mathbm{G}_{\mathrm{T},k} \mathbm{M}_{t,k}^{-1} \mathbm{G}_{\mathrm{T},k}^H \mathbm{x}_t - 2 \mu ( \widetilde{\mathbm{M}}_{t,k} - \ell_{\text{max},k} \mathbm{I}_{N_{\mathrm{T}}} ) \mathbm{x}_t ,
\end{equation}
and 
\begin{equation}
	\begin{aligned}
		\text{const}_{2,k} = & 
		- \mu \text{Tr} \{ \mathbm{M}_{t,k}^{-1} \mathbm{G}_{\mathrm{T},k} \mathbm{x}_t \mathbm{x}_t^H \mathbm{G}_{\mathrm{T},k}^H \mathbm{M}_{t,k}^{-1} \} \\
		& - \mu \ell_{\text{max},k}E + \mu \mathbm{x}_t ( \widetilde{\mathbm{M}}_{t,k} - \ell_{\text{max},k} \mathbm{I}_{N_{\mathrm{T}}} ) \mathbm{x}_t  ,
	\end{aligned}
\end{equation}
with 
$\mathbm{M}_{t,k} = \sum_{p = 1, p\ne k}^K \mathrm{SNR}_{\mathrm{R},p} \mathbm{G}_{\mathrm{T}, p} \mathbm{x} \mathbm{x}^H \mathbm{G}_{\mathrm{T},p}
+ \sum_{q = 1}^ Q \mathrm{CNR}_{\mathrm{R},q} \mathbm{G}_q \mathbm{x} \mathbm{x}^H \mathbm{G}_q^H + \mathbm{I}_{N_{\mathrm{R}}}$,
$\widetilde{\mathbm{M}}_{t,k}
= \sum_{p=1, p \ne k}^K \mathrm{SNR}_{\mathrm{R},p} {\mathbm{G}}_{\mathrm{T},p}^H \mathbm{M}_{t,k}^{-1} \mathbm{G}_{\mathrm{T},k} \mathbm{x}_t \mathbm{x}_t^H \mathbm{G}_{\mathrm{T},k}^H \mathbm{M}_{t,k}^{-1} {\mathbm{G}}_{\mathrm{T},p}
+ \sum_{q=1}^Q \mathrm{CNR}_{\mathrm{R},q} {\mathbm{G}}_q^H \mathbm{M}_{t,k}^{-1} \mathbm{G}_{\mathrm{T},k} \mathbm{x}_t \mathbm{x}_t^H \mathbm{G}_{\mathrm{T},k}^H \mathbm{M}_{t,k}^{-1} {\mathbm{G}}_q$,
and $\ell_{\text{max},k}$ beeing maximum eigenvalue of $\widetilde{\mathbm{M}}_{t,k}$.

Based on above transformations, we can equivalently reformulate \eqref{eq_57} as
\begin{subequations}
	\begin{align}
		\mathop {\max }\limits_{{\mathbm{x}},\{\mathbm{f}_k\}, \rho} \quad & \rho \\
		{\text{s}}.{\text{t}}. \;\quad & \Re \{ \mathbm{Ax} \} \ge \mathbm{\lambda} , \\
		& \Re\left\{ \mathbm{w}_{t,k}^H \mathbm{x} \right\} + \text{const}_{2,k} \ge \rho , \forall k , \\
		& {\mathbm{x}} \in \mathcal{X}_{\text{DAC}}^{\text{1-Bit}} = \left\{ \frac{E}{{\sqrt {2{M_{\text{T}}}} }}  ({ \pm 1 \pm \jmath}) \right\} ,
	\end{align}
\end{subequations}
which can be further transformed into its real-valued form (See Sec. \ref{Sec_4_B_3}) and solved via ILP solvers.

\footnotesize
\bibliographystyle{IEEEtran}
\bibliography{IEEEabrv,./ref.bib}

\begin{thebibliography}{10}
\providecommand{\url}[1]{#1}
\csname url@samestyle\endcsname
\providecommand{\newblock}{\relax}
\providecommand{\bibinfo}[2]{#2}
\providecommand{\BIBentrySTDinterwordspacing}{\spaceskip=0pt\relax}
\providecommand{\BIBentryALTinterwordstretchfactor}{4}
\providecommand{\BIBentryALTinterwordspacing}{\spaceskip=\fontdimen2\font plus
\BIBentryALTinterwordstretchfactor\fontdimen3\font minus
  \fontdimen4\font\relax}
\providecommand{\BIBforeignlanguage}[2]{{%
\expandafter\ifx\csname l@#1\endcsname\relax
\typeout{** WARNING: IEEEtran.bst: No hyphenation pattern has been}%
\typeout{** loaded for the language `#1'. Using the pattern for}%
\typeout{** the default language instead.}%
\else
\language=\csname l@#1\endcsname
\fi
#2}}
\providecommand{\BIBdecl}{\relax}
\BIBdecl

\bibitem{wang2024massive_arXiv}
B.~Wang, H.~Li, B.~Liao, and Z.~Cheng, ``Massive {MIMO-ISAC} system with 1-bit
  {ADCs/DACs},'' \emph{arXiv preprint arXiv:2405.15553}, 2024.

\bibitem{wang2023joint}
B.~Wang, H.~Li, and Z.~Cheng, ``Joint transceiver design for massive {MIMO}
  {DFRC} systems with one-bit {DACs/ADCs},'' in \emph{Proc. of 2023 IEEE
  Globecom Workshops (GC Wkshps)}.\hskip 1em plus 0.5em minus 0.4em\relax IEEE,
  2023, pp. 649--654.

\bibitem{liu2023seventy}
F.~Liu, L.~Zheng, Y.~Cui, C.~Masouros, A.~P. Petropulu, H.~Griffiths, and Y.~C.
  Eldar, ``Seventy years of radar and communications: The road from separation
  to integration,'' \emph{{IEEE} Signal Process. Mag.}, vol.~40, no.~5, pp.
  106--121, 2023.

\bibitem{zhang2021overview}
J.~A. Zhang, F.~Liu, C.~Masouros, R.~W. Heath, Z.~Feng, L.~Zheng, and
  A.~Petropulu, ``An overview of signal processing techniques for joint
  communication and radar sensing,'' \emph{{IEEE} J. Sel. Topics Signal
  Process.}, vol.~15, no.~6, pp. 1295--1315, 2021.

\bibitem{liu2020joint}
F.~Liu, C.~Masouros, A.~P. Petropulu, H.~Griffiths, and L.~Hanzo, ``Joint radar
  and communication design: Applications, state-of-the-art, and the road
  ahead,'' \emph{{IEEE} Trans. Commun.}, vol.~68, no.~6, pp. 3834--3862, 2020.

\bibitem{xiong2024torch}
Y.~Xiong, F.~Liu, K.~Wan, W.~Yuan, Y.~Cui, and G.~Caire, ``From torch to
  projector: Fundamental tradeoff of integrated sensing and communications,''
  \emph{IEEE BITS the Inform. Theory Mag.}, vol.~4, no.~1, pp. 73--90, 2024.

\bibitem{feng2011received}
C.~Feng, W.~S.~A. Au, S.~Valaee, and Z.~Tan, ``Received-signal-strength-based
  indoor positioning using compressive sensing,'' \emph{{IEEE} Trans. Mobile
  Comput.}, vol.~11, no.~12, pp. 1983--1993, 2011.

\bibitem{jiang2008ieee}
D.~Jiang and L.~Delgrossi, ``{IEEE 802.11 p}: Towards an international standard
  for wireless access in vehicular environments,'' in \emph{Proc. of 2008 IEEE
  veh. technol. conf. (VTC Spring)}.\hskip 1em plus 0.5em minus 0.4em\relax
  IEEE, 2008, pp. 2036--2040.

\bibitem{kumari2017ieee}
P.~Kumari, J.~Choi, N.~Gonz{\'a}lez-Prelcic, and R.~W. Heath, ``{IEEE 802.11}
  ad-based radar: An approach to joint vehicular communication-radar system,''
  \emph{{IEEE} Trans. Veh. Technol.}, vol.~67, no.~4, pp. 3012--3027, 2017.

\bibitem{li2024frame}
Y.~Li, F.~Liu, Z.~Du, W.~Yuan, Q.~Shi, and C.~Masouros, ``Frame structure and
  protocol design for sensing-assisted {NR-V2X} communications,'' \emph{{IEEE}
  Trans. Mobile Comput.}, vol.~23, no.~12, pp. 11\,045--11\,060, 2024.

\bibitem{nowak2016co}
M.~Nowak, M.~Wicks, Z.~Zhang, and Z.~Wu, ``Co-designed radar-communication
  using linear frequency modulation waveform,'' \emph{{IEEE} Aerosp. Electron.
  Syst. Mag.}, vol.~31, no.~10, pp. 28--35, 2016.

\bibitem{wu2021frequency}
K.~Wu, J.~A. Zhang, X.~Huang, and Y.~J. Guo, ``Frequency-hopping {MIMO}
  radar-based communications: An overview,'' \emph{{IEEE} Aerosp. Electron.
  Syst. Mag.}, vol.~37, no.~4, pp. 42--54, 2021.

\bibitem{ahmed2018dual}
A.~Ahmed, Y.~D. Zhang, and Y.~Gu, ``Dual-function radar-communications using
  {QAM}-based sidelobe modulation,'' \emph{Digital Signal Process.}, vol.~82,
  pp. 166--174, 2018.

\bibitem{meng2023network}
K.~Meng, C.~Masouros, G.~Chen, and F.~Liu, ``Network-level integrated sensing
  and communication: Interference management and {BS} coordination using
  stochastic geometry,'' \emph{{IEEE} Trans. Wireless Commun.}, vol.~23,
  no.~12, pp. 19\,365--19\,381, 2024.

\bibitem{liao2024faster}
Z.~Liao, F.~Liu, A.~Li, and C.~Masouros, ``Faster-than-nyquist symbol-level
  precoding for wideband integrated sensing and communications,'' \emph{{IEEE}
  Trans. Wireless Commun.}, vol.~23, no.~8, pp. 10\,445--10\,458, 2024.

\bibitem{liu2020TSP}
X.~Liu, T.~Huang, N.~Shlezinger, Y.~Liu, J.~Zhou, and Y.~C. Eldar, ``Joint
  transmit beamforming for multiuser {MIMO} communications and {MIMO} radar,''
  \emph{{IEEE} Trans. Signal Process.}, vol.~68, pp. 3929--3944, 2020.

\bibitem{liu2021dual}
R.~Liu, M.~Li, Q.~Liu, and A.~L. Swindlehurst, ``Dual-functional
  radar-communication waveform design: A symbol-level precoding approach,''
  \emph{{IEEE} J. Sel. Topics Signal Process.}, vol.~15, no.~6, pp. 1316--1331,
  2021.

\bibitem{liu2021cramer}
F.~Liu, Y.-F. Liu, A.~Li, C.~Masouros, and Y.~C. Eldar, ``{Cram{\'e}r-{Rao}}
  bound optimization for joint radar-communication beamforming,'' \emph{{IEEE}
  Trans. Signal Process.}, vol.~70, pp. 240--253, 2021.

\bibitem{Guo2023TSP}
B.~Guo, J.~Liang, B.~Tang, L.~Li, and H.~C. So, ``Bistatic {MIMO DFRC} system
  waveform design via symbol distance/direction discrimination,'' \emph{{IEEE}
  Trans. Signal Process.}, vol.~71, pp. 3996--4010, 2023.

\bibitem{wei2023waveform}
Z.~Wei, J.~Piao, X.~Yuan, H.~Wu, J.~A. Zhang, Z.~Feng, L.~Wang, and P.~Zhang,
  ``Waveform design for {MIMO-OFDM} integrated sensing and communication
  system: An information theoretical approach,'' \emph{{IEEE} Trans. Commun.},
  vol.~72, no.~1, pp. 496--509, 2024.

\bibitem{wang2023relative}
X.~Wang, B.~Tang \emph{et~al.}, ``Relative entropy-based waveform optimization
  for rician target detection with dual-function radar communication systems,''
  \emph{{IEEE} Sensors J.}, vol.~23, no.~10, pp. 10\,718--10\,730, 2023.

\bibitem{Sun2025JSAC}
Y.~Sun, K.~An, M.~Yu, Y.~Hu, Y.~Zhu, Z.~Lin, M.~Xiao, N.~Al-Dhahir, D.~Niyato,
  and J.~Wang, ``{Dual-Polarized Stacked Metasurface Transceiver Design With
  Rate Splitting for Next-Generation Wireless Networks},'' \emph{{IEEE} J. Sel.
  Areas Commun.}, vol.~43, no.~3, pp. 811--833, 2025.

\bibitem{Sun2024TWC}
Y.~Sun, Y.~Zhu, K.~An, Z.~Lin, C.~Li, D.~W.~K. Ng, and J.~Wang,
  ``{Active-Passive Cascaded {RIS}-Aided Receiver Design for Jamming Nulling
  and Signal Enhancing},'' \emph{{IEEE} Trans. Wireless Commun.}, vol.~23,
  no.~6, pp. 5345--5362, 2024.

\bibitem{zhang2018low}
J.~Zhang, L.~Dai, X.~Li, Y.~Liu, and L.~Hanzo, ``On low-resolution {ADCs} in
  practical {5G} millimeter-wave massive {MIMO} systems,'' \emph{{IEEE} Commun.
  Mag.}, vol.~56, no.~7, pp. 205--211, 2018.

\bibitem{castaneda20171}
O.~Casta{\~n}eda \emph{et~al.}, ``1-bit massive {MU-MIMO} precoding in
  {VLSI},'' \emph{IEEE J. on Emerging and Sel. Topics in Circuits and Syst.},
  vol.~7, no.~4, pp. 508--522, 2017.

\bibitem{li20211}
A.~Li, C.~Masouros, A.~L. Swindlehurst, and W.~Yu, ``1-bit massive {MIMO}
  transmission: Embracing interference with symbol-level precoding,''
  \emph{{IEEE} Commun. Mag.}, vol.~59, no.~5, pp. 121--127, 2021.

\bibitem{mo2017channel}
J.~Mo, P.~Schniter, and R.~W. Heath, ``Channel estimation in broadband
  millimeter wave {MIMO} systems with few-bit {ADCs},'' \emph{{IEEE} Trans.
  Signal Process.}, vol.~66, no.~5, pp. 1141--1154, 2017.

\bibitem{Shao2019TSP}
M.~Shao \emph{et~al.}, ``A framework for one-bit and constant-envelope
  precoding over multiuser massive {MISO} channels,'' \emph{{IEEE} Trans.
  Signal Process.}, vol.~67, no.~20, pp. 5309--5324, 2019.

\bibitem{Wu2023CI}
Z.~Wu, B.~Jiang, Y.-F. Liu, M.~Shao, and Y.-H. Dai, ``Efficient {CI}-based
  one-bit precoding for multiuser downlink massive {MIMO} systems with {PSK}
  modulation,'' \emph{{IEEE} Trans. Wireless Commun.}, vol.~23, no.~5, pp.
  4861--4875, 2024.

\bibitem{meng2018generalized}
X.~Meng and J.~Zhu, ``A generalized sparse bayesian learning algorithm for
  1-bit {DOA} estimation,'' \emph{{IEEE} Commun. Lett.}, vol.~22, no.~7, pp.
  1414--1417, 2018.

\bibitem{Nghi2020TWC}
M.~Ranjbar, N.~H. Tran, M.~N. Vu, T.~V. Nguyen, and M.~Cenk~Gursoy, ``Capacity
  region and capacity-achieving signaling schemes for 1-bit adc multiple access
  channels in rayleigh fading,'' \emph{{IEEE} Trans. Wireless Commun.},
  vol.~19, no.~9, pp. 6162--6178, 2020.

\bibitem{Nghi2025Physical}
M.~H. Rahman, H.~Nguyen-Le, N.~H. Tran, H.~V. Nguyen, M.~T. Le, and Y.~Zhang,
  ``Boundedness of optimal input signals for two-user multiple access channels
  with 1-bit quantization and gaussian mixture noise with arbitrary means and
  variances,'' \emph{Physical Communication}, p. 102778, 2025.

\bibitem{cheng2023relative}
Z.~Cheng, L.~Wu, B.~Wang, J.~Xie, and H.~Li, ``Relative entropy-based
  constant-envelope beamforming for target detection in large-scale {MIMO}
  radar with low-resoultion {ADCs},'' \emph{{IEEE} Trans. Veh. Technol.},
  vol.~72, no.~8, pp. 10\,090--10\,106, 2023.

\bibitem{xi2020gridless}
F.~Xi, Y.~Xiang, S.~Chen, and A.~Nehorai, ``Gridless parameter estimation for
  one-bit {MIMO} radar with time-varying thresholds,'' \emph{{IEEE} Trans.
  Signal Process.}, vol.~68, pp. 1048--1063, 2020.

\bibitem{xiao2022one}
Y.-H. Xiao, D.~Ram{\'\i}rez, P.~J. Schreier \emph{et~al.}, ``One-bit target
  detection in collocated {MIMO} radar and performance degradation analysis,''
  \emph{{IEEE} Trans. Veh. Technol.}, vol.~71, no.~9, pp. 9363--9374, 2022.

\bibitem{deng2022receive}
M.~Deng \emph{et~al.}, ``One-bit {ADCs/DACs} based {MIMO} radar: Performance
  analysis and joint design,'' \emph{{IEEE} Trans. Signal Process.}, vol.~70,
  pp. 2609--2624, 2022.

\bibitem{shang2024mixed}
X.~Shang, R.~Lin, and Y.~Cheng, ``Mixed-{ADC} based {PMCW MIMO} radar
  angle-{Doppler} imaging,'' \emph{{IEEE} Trans. Signal Process.}, vol.~72, pp.
  883--895, 2024.

\bibitem{Liao2019MUSIC}
X.~Huang and B.~Liao, ``One-bit {MUSIC},'' \emph{{IEEE} Signal Process. Lett.},
  vol.~26, no.~7, pp. 961--965, 2019.

\bibitem{Sedighi2021TSP}
S.~Sedighi, B.~S. Mysore~R, M.~Soltanalian, and B.~Ottersten, ``On the
  performance of one-bit {DoA} estimation via sparse linear arrays,''
  \emph{{IEEE} Trans. Signal Process.}, vol.~69, pp. 6165--6182, 2021.

\bibitem{cheng2021transmit}
Z.~Cheng, S.~Shi, Z.~He, and B.~Liao, ``Transmit sequence design for
  dual-function radar-communication system with one-bit {DACs},'' \emph{{IEEE}
  Trans. Wireless Commun.}, vol.~20, no.~9, pp. 5846--5860, 2021.

\bibitem{yu2022precoding}
X.~Yu, Q.~Yang, Z.~Xiao, H.~Chen, V.~Havyarimana, and Z.~Han, ``A precoding
  approach for dual-functional radar-communication system with one-bit
  {DACs},'' \emph{{IEEE} J. Sel. Areas Commun.}, vol.~40, no.~6, pp.
  1965--1977, 2022.

\bibitem{Lin2025TCOM}
Q.~Lin, H.~Shen, Z.~Li, W.~Xu, C.~Zhao, and X.~You, ``One-bit transceiver
  optimization for mmwave integrated sensing and communication systems,''
  \emph{{IEEE} Trans. Commun.}, vol.~73, no.~2, pp. 800--816, 2025.

\bibitem{papoulis1965random}
A.~Papoulis, \emph{Random variables and stochastic processes}.\hskip 1em plus
  0.5em minus 0.4em\relax McGraw Hill, 1965.

\bibitem{li2013distributed}
M.~Li, C.~Liu, and S.~V. Hanly, ``Distributed base station cooperation with
  finite alphabet and {QoS} constraints,'' in \emph{Proc. IEEE Int. Symp. Inf.
  Theory (ISIT), 2013}.\hskip 1em plus 0.5em minus 0.4em\relax IEEE, 2013, pp.
  1157--1161.

\bibitem{salem2021error}
A.~Salem and C.~Masouros, ``Error probability analysis and power allocation for
  interference exploitation over rayleigh fading channels,'' \emph{{IEEE}
  Trans. Wireless Commun.}, vol.~20, no.~9, pp. 5754--5768, 2021.

\bibitem{shao2019framework}
M.~Shao, Q.~Li, W.-K. Ma, and A.~M.-C. So, ``A framework for one-bit and
  constant-envelope precoding over multiuser massive {MISO} channels,''
  \emph{{IEEE} Trans. Signal Process.}, vol.~67, no.~20, pp. 5309--5324, 2019.

\bibitem{li2020interference}
A.~Li, F.~Liu, C.~Masouros, Y.~Li, and B.~Vucetic, ``Interference exploitation
  1-bit massive {MIMO} precoding: A partial branch-and-bound solution with
  near-optimal performance,'' \emph{{IEEE} Trans. Wireless Commun.}, vol.~19,
  no.~5, pp. 3474--3489, 2020.

\bibitem{li2018massive}
A.~Li, C.~Masouros, F.~Liu, and A.~L. Swindlehurst, ``Massive {MIMO} 1-bit
  {DAC} transmission: A low-complexity symbol scaling approach,'' \emph{{IEEE}
  Trans. Wireless Commun.}, vol.~17, no.~11, pp. 7559--7575, 2018.

\bibitem{jacobsson2017quantized}
S.~Jacobsson, G.~Durisi, M.~Coldrey, T.~Goldstein, and C.~Studer, ``Quantized
  precoding for massive {MU-MIMO},'' \emph{{IEEE} Trans. Commun.}, vol.~65,
  no.~11, pp. 4670--4684, 2017.

\bibitem{Chen2021Pareto}
L.~Chen, F.~Liu, W.~Wang, and C.~Masouros, ``Joint radar-communication
  transmission: A generalized {Pareto} optimization framework,'' \emph{{IEEE}
  Trans. Signal Process.}, vol.~69, pp. 2752--2765, 2021.

\bibitem{cui2013mimo}
G.~Cui, H.~Li, and M.~Rangaswamy, ``{MIMO} radar waveform design with constant
  modulus and similarity constraints,'' \emph{{IEEE} Trans. Signal Process.},
  vol.~62, no.~2, pp. 343--353, 2013.

\bibitem{stein2013quantization}
M.~Stein, F.~Wendler, A.~Mezghani, and J.~A. Nossek, ``Quantization-loss
  reduction for signal parameter estimation,'' in \emph{2013 Proc. IEEE Int.
  Conf. Acoust., Speech, Signal Process. (ICASSP)}.\hskip 1em plus 0.5em minus
  0.4em\relax IEEE, 2013, pp. 5800--5804.

\bibitem{mo2017hybrid}
J.~Mo \emph{et~al.}, ``Hybrid architectures with few-bit {ADC} receivers:
  Achievable rates and energy-rate tradeoffs,'' \emph{{IEEE} Trans. Wireless
  Commun.}, vol.~16, no.~4, pp. 2274--2287, 2017.

\bibitem{roth2017achievable}
K.~Roth and J.~A. Nossek, ``Achievable rate and energy efficiency of hybrid and
  digital beamforming receivers with low resolution {ADC},'' \emph{IEEE Journal
  on Selected Areas in Communications}, vol.~35, no.~9, pp. 2056--2068, 2017.

\bibitem{simon2004digital}
M.~K. Simon and M.-S. Alouini, \emph{Digital communication over fading
  channels}.\hskip 1em plus 0.5em minus 0.4em\relax John Wiley \& Sons, 2004.

\bibitem{shao2018multiuser}
M.~Shao, Q.~Li, Y.~Liu, and W.-K. Ma, ``Multiuser one-bit massive {MIMO}
  precoding under {MPSK} signaling,'' in \emph{2018 IEEE Global Conference on
  Signal and Information Processing (GlobalSIP)}.\hskip 1em plus 0.5em minus
  0.4em\relax IEEE, 2018, pp. 833--837.

\bibitem{Wu2023Diversity}
Z.~Wu, J.~Wu, W.-K. Chen, and Y.-F. Liu, ``Diversity order analysis for
  quantized constant envelope transmission,'' \emph{IEEE Open J. Signal
  Process.}, vol.~4, pp. 21--30, 2023.

\bibitem{sun2016majorization}
Y.~Sun, P.~Babu, and D.~P. Palomar, ``Majorization-minimization algorithms in
  signal processing, communications, and machine learning,'' \emph{{IEEE}
  Trans. Signal Process.}, vol.~65, no.~3, pp. 794--816, 2016.

\bibitem{landau2017branch}
L.~T. Landau and R.~C. de~Lamare, ``Branch-and-bound precoding for multiuser
  {MIMO} systems with 1-bit quantization,'' \emph{{IEEE} Wireless Commun.
  Lett.}, vol.~6, no.~6, pp. 770--773, 2017.

\bibitem{boyd2007branch}
S.~Boyd and J.~Mattingley, ``Branch and bound methods,'' \emph{Notes for
  EE364b, Stanford University}, vol. 2006, p.~07, 2007.

\bibitem{das2024branch}
S.~Das~Gupta, B.~P. Van~Parys, and E.~K. Ryu, ``Branch-and-bound performance
  estimation programming: A unified methodology for constructing optimal
  optimization methods,'' \emph{Mathematical Programming}, vol. 204, no.~1, pp.
  567--639, 2024.

\bibitem{walden1999analog}
R.~H. Walden, ``Analog-to-digital converter survey and analysis,'' \emph{{IEEE}
  J. Sel. Areas Commun.}, vol.~17, no.~4, pp. 539--550, 1999.

\bibitem{STCOM2026}
Y.~Sun, R.~Liu, M.~Li, and Q.~Liu, ``1-bit {DAC/ADC} transceiver designs for
  efficient mimo-isac systems,'' \emph{{IEEE} Trans. Commun.}, vol.~74, pp.
  4694--4709, 2026.

\bibitem{billingsley2013convergence}
P.~Billingsley, \emph{Convergence of probability measures}.\hskip 1em plus
  0.5em minus 0.4em\relax John Wiley \& Sons, 2013.

\bibitem{williams1991probability}
D.~Williams, \emph{Probability with martingales}.\hskip 1em plus 0.5em minus
  0.4em\relax Cambridge university press, 1991.

\bibitem{billingsley2008probability}
P.~Billingsley, \emph{Probability and measure}.\hskip 1em plus 0.5em minus
  0.4em\relax John Wiley \& Sons, 2008.

\end{thebibliography}

\end{document}